\documentclass[twocolumn]{aastex61}
\usepackage{graphicx,subfigure,amsmath, amsfonts, amssymb,footnote,color,epstopdf,ulem}
\usepackage[bookmarks=false]{hyperref}
\usepackage{apjfonts}
\hypersetup{
  colorlinks= true, 
  urlcolor  = blue, 
  linkcolor = red, 
  citecolor = blue 
} 
\def\hi{H{\small I}}

\shorttitle{AGNs in the ngVLA Era}
\shortauthors{Kristina Nyland et al.}

\begin{document}

\title{Revolutionizing Our Understanding of AGN Feedback and its Importance to Galaxy Evolution in the Era of the Next Generation Very Large Array}

\correspondingauthor{Kristina Nyland}
\email{knyland@nrao.edu}

\author{K. Nyland}
\affiliation{National Radio Astronomy Observatory, Charlottesville, VA 22903, USA}

\author{J.~J. Harwood}
\affiliation{ASTRON, The Netherlands Institute for Radio Astronomy, Postbus 2, NL-7990 AA Dwingeloo, the Netherlands}

\author{D.~D. Mukherjee}
\affiliation{Dipartimento di Fisica Generale, Universita degli Studi di Torino, Via Pietro Giuria 1, 10125 Torino, Italy}

\author{P. Jagannathan}
\affiliation{National Radio Astronomy Observatory, Socorro, NM 87801, USA}

\author{W. Rujopakarn}
\affiliation{Department of Physics, Faculty of Science, Chulalongkorn University, 254 Phayathai Road, Pathumwan, Bangkok 10330, Thailand}
\affiliation{National Astronomical Research Institute of Thailand (Public Organization), Don Kaeo, Mae Rim, Chiang Mai 50180, Thailand}
\affiliation{Kavli Institute for the Physics and Mathematics of the universe (WPI), The University of Tokyo Institutes for Advanced Study, The University of Tokyo, Kashiwa, Chiba 277-8583, Japan}

\author{B. Emonts}
\affiliation{National Radio Astronomy Observatory, Charlottesville, VA 22903, USA}

\author{K. Alatalo}\altaffiliation{Hubble fellow}
\affiliation{Space Telescope Science Institute, 3700 San Martin Dr., Baltimore, MD 21218, USA}
\affiliation{The Observatories of the Carnegie Institution for Science, 813 Santa Barbara St., Pasadena, CA 91101, USA}

\author{G.~V. Bicknell}
\affiliation{Research School of Astronomy and Astrophysics, Australian National University, Canberra, ACT 2611, Australia}

\author{T.~A. Davis}
\affiliation{School of Physics \&\ Astronomy, Cardiff University, Queens Buildings, The Parade, Cardiff, CF24 3AA, UK}

\author{J.~E. Greene}
\affiliation{Department of Astrophysics, Princeton University, Princeton, NJ 08540, USA}

\author{A. Kimball}
\affiliation{National Radio Astronomy Observatory, 1003 Lopezville Rd., Socorro, NM, 87801, USA}

\author{M. Lacy}
\affiliation{National Radio Astronomy Observatory, Charlottesville, VA 22903, USA}

\author{Carol Lonsdale}
\affiliation{National Radio Astronomy Observatory, Charlottesville, VA 22903, USA}

\author{Colin Lonsdale}
\affiliation{MIT Haystack Observatory, Westford, MA 01886, USA}

\author{W.~P. Maksym}
\affiliation{Harvard-Smithsonian Center for Astrophysics, 60 Garden St., Cambridge, MA, 02478, USA}

\author{D.~C. {Moln{\'a}r}}
\affiliation{Astronomy Centre, Department of Physics and Astronomy, University of Sussex, Brighton BN1 9QH, UK}

\author{L. Morabito}
\affiliation{Astrophysics, University of Oxford, Denys Wilkinson Building, Keble Road, Oxford, OX1 3RH}

\author{E.~J. Murphy}
\affiliation{National Radio Astronomy Observatory, Charlottesville, VA 22903, USA}

\author{P. Patil}
\affiliation{National Radio Astronomy Observatory, Charlottesville, VA 22903, USA}
\affiliation{University of Virginia, Department of Astronomy, Charlottesville, VA 22903, USA}

\author{I. Prandoni}
\affiliation{INAF-Istituto di Radioastronomia, via P. Gobetti 101, 40129, Bologna, Italy}

\author{M. Sargent}
\affiliation{Astronomy Centre, Department of Physics and Astronomy, University of Sussex, Brighton BN1 9QH, UK}

\author{C. Vlahakis}
\affiliation{National Radio Astronomy Observatory, Charlottesville, VA 22903, USA}

\begin{abstract}
Energetic feedback by Active Galactic Nuclei (AGNs) plays an important evolutionary role in the regulation of star formation (SF) on galactic scales. However, the effects of this feedback as a function of redshift and galaxy properties such as mass, environment and cold gas content remain poorly understood.  The broad frequency coverage (1 to 116 GHz), high sensitivity (up to ten times higher than the Karl G. Jansky Very Large Array), and superb angular resolution (maximum baselines of at least a few hundred km) of the proposed next generation Very Large Array (ngVLA) are uniquely poised to revolutionize our understanding of AGNs and their role in galaxy evolution.  Here, we provide an overview of the science related to AGN feedback that will be possible in the ngVLA era and present new continuum ngVLA imaging simulations of resolved radio jets spanning a wide range of intrinsic extents.  We also consider key computational challenges and discuss exciting opportunities for multi-wavelength synergy with other next-generation instruments, such as the Square Kilometer Array and the {\it James Webb} Space Telescope.  The unique combination of high-resolution, large collecting area, and wide frequency range will enable significant advancements in our understanding of the effects of jet-driven feedback on sub-galactic scales, particularly for sources with extents of a few pc to a few kpc such as young and/or lower-power radio AGNs, AGNs hosted by low-mass galaxies, radio jets that are interacting strongly with the interstellar medium of the host galaxy, and AGNs at high redshift.
\end{abstract}


\section{Introduction and Motivation}
Decades of observations of active galactic nuclei (AGNs) and their host galaxies have provided strong evidence that energetic AGN-driven feedback plays a pivotal role in influencing galaxy evolution.  AGNs have been implicated in establishing the scaling relations between massive black hole (MBH) and host galaxy properties, regulating cooling flows in clusters, driving galaxy-scale outflows, and contributing to the build-up of the red sequence of massive galaxies \citep{kormendy+13, heckman+14}.  Cosmological simulations have provided further support for the importance of this paradigm by demonstrating the inability of models {\it lacking} energetic AGN feedback to produce the observed distribution of galaxy masses at $z = 0$ (e.g., \citealt{kaviraj+17}).  

The importance of AGN feedback in the context of galaxy evolution stems from the regulatory effect it may have on the star formation rate and efficiency of the host galaxy. AGN feedback may be driven by radiative winds launched by the accretion disks of powerful quasars (``radiative'' or 	``quasar'' mode feedback) or spurred by radio jets/lobes as they heat, expel, or shock their surroundings (``radio'' or ``jet'' mode feedback).  Observational evidence for both modes of feedback has been reported (e.g., \citealt{fabian+12}, and references therein). \citet{villar-martin+14} show that, on average, radio jets appear to be capable of producing more extreme gas outflows than accretion disk winds. However, the relative importance of each mode, and the dependence on redshift, remains an open area of research.  Here, we focus on opportunities for improving our understanding of jet-driven AGN feedback using future radio telescopes.  

Despite the importance of radio-mode AGN feedback, our understanding of the underlying {\it physics} of jetted AGNs, as well as their evolutionary role in spiral and lower-mass galaxies, remains shockingly incomplete.  We still lack a fundamental understanding of exactly {\it how} radio jets transfer energy to their surroundings, and to what degree this deposition of energy affects the evolution of different types of galaxies over cosmic time.  Progress in our understanding of the physics of radio AGNs and their role in galaxy evolution will require deep continuum and spectral line radio observations over an unprecedented range of frequencies and spatial scales.  

The fundamental limitations of existing radio telescopes that are currently preventing substantial progress in this area include sensitivity, angular resolution, and frequency range.  The most significant of these limitations is sensitivity.  Bright radio AGNs, which include radio quasars and classical radio galaxies with jets/lobes extending far beyond the optical extents of their host galaxies, comprise only a small fraction of the AGN population.  The majority of AGNs ($\sim$~90\%; \citealt{padovani+16}, and references therein) are characterized by radio luminosities of $L_{1.4\,\mathrm{GHz}} \lesssim  10^{24}$ W~Hz$^{-1}$ and are considered to be in the ``radio quiet'' (but not radio silent) category\footnote{We refer to this class of AGNs, which may be either jetted/non-jetted, interchangeably as radio quiet, low-luminosity, and low-power radio AGNs throughout this study.}.  This class of objects consists of a mix of jetted and non-jetted AGNs\footnote{Some radio sources with $L_{1.4\,\mathrm{GHz}} \lesssim  10^{24}$ W~Hz$^{-1}$ have star formation rather than AGN origins, although the prevalence of this phenomenon is widely debated in the literature and likely varies as a function of redshift, galaxy properties, and survey selection effects (e.g., \citealt{kimball+11, white+15, zakamska+16}).} that are typically characterized by inefficient MBH accretion with Eddington ratios below $\sim$1\%, though higher-accretion-rate jetted AGNs, such as Seyferts and high-excitation radio galaxies, also contribute to this population  \citep{ho+08, heckman+14, padovani+16}.  Although this population of intrinsically faint radio AGNs far outnumbers luminous radio galaxies, their properties and role in galaxy evolution remain poorly understood due to the sensitivity limitations of current instruments such as the Karl G. Jansky Very Large Array (VLA).

Angular resolution is also of paramount importance for improving our understanding of AGNs and their connection to galaxy evolution.  Radio continuum imaging at high spatial resolution facilitates accurate cross-identification of faint, high-redshift radio sources with their hosts in crowded optical/infrared images (e.g., \citealt{rujopakarn+16, murphy+17}).  Furthermore, high-resolution imaging that is sensitive to emission over a wide range of spatial scales is necessary for unambiguously identifying radio AGNs based on their morphologies.  This is of particular importance for studies of the dominant population of lower-power AGNs, which tend to be faint at radio wavelengths, and therefore require high-resolution imaging to distinguish between the signatures of MBH accretion and stellar processes (e.g., \citealt{nyland+17}).  High-resolution imaging is also necessary for measuring the physical extents of radio jets/lobes needed for energetics estimates, determining the scope of jet-driven feedback (circumnuclear vs.\ galactic-scale feedback), and comparing observations of radio jets interacting with their hosts with simulations.

Another critical aspect of quantifying the physics of jet-driven feedback and placing it in the context of galaxy evolution is constraining the ages of radio AGNs through broadband continuum imaging.  The ages of jetted AGNs may be estimated by comparing continuum observations over a wide range of frequencies with radio source spectral aging models (e.g., \citealt{myers+85, carilli+91, harwood+13}).  In addition, a sufficiently wide frequency range encompassing not only the 21~cm neutral hydrogen (\hi) line but also molecular gas emission from the CO(1--0) transition at 115.2712~GHz would provide measurements of the interstellar medium (ISM) content and conditions in the immediate vicinity of AGNs.  Ultimately, this would provide direct observational constraints on the energetic impact of radio jets hosted by gas-rich galaxies as the jets interact with the ambient ISM, which could then be compared with cutting-edge hydrodynamic simulations of radio jets propagating through a dense medium (e.g., \citealt{mukherjee+16, mukherjee+17}).  

In this study, we evaluate the progress in our understanding of AGN feedback and its connection to galaxy evolution that could be accomplished with the unique capabilities of the Next Generation Very Large Array (ngVLA; \citealt{murphy+17b}), a prospective new radio telescope for the 2030's that is currently in the early stages of design and development by the National Radio Astronomy Observatory (NRAO).  The goal of the ngVLA project is to design an instrument with up to ten times higher sensitivity and spatial resolution compared to the VLA that will operate over a broad frequency range spanning 1 to 116~GHz.  Thus, the ngVLA will be sensitive to continuum emission over a wide range of frequencies as well as to cold gas traced by the H{\tt I} line at 21~cm and low-$J$ transitions of the CO molecule. 

\begin{figure}
\centering
\includegraphics[clip=true, trim=6.5cm 1.25cm 5cm 1.25cm, height=0.425\textwidth]{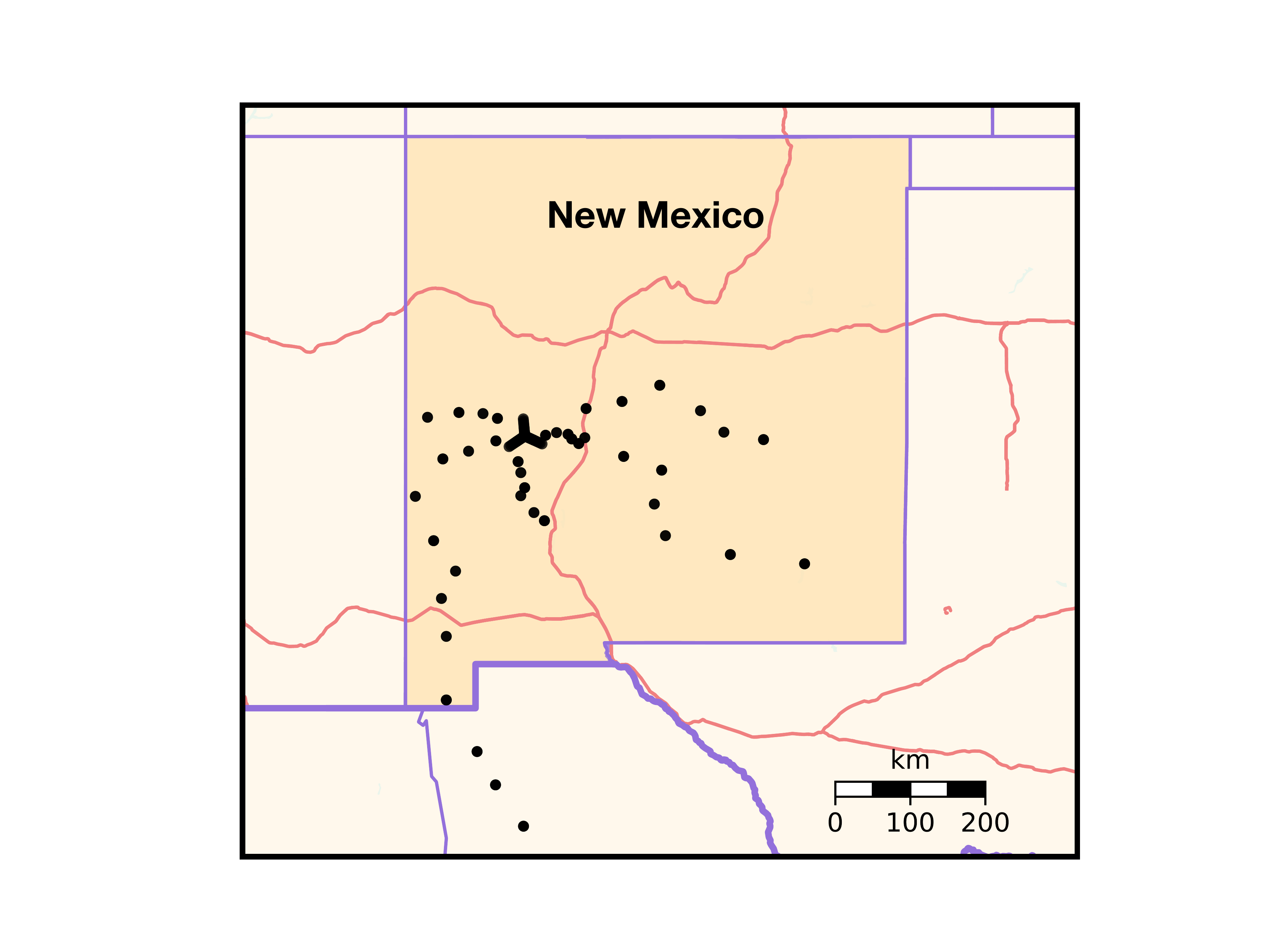}
\caption{Antenna locations in the ngVLA configuration considered in this study.  Major U.S. highways are shown in red. The array is centered on the compact core of the current VLA in New Mexico, and the additional antennas at intermediate and long baselines take into account local geography, land ownership, and infrastructure.  The longest baselines of 513~km extend in the north-south direction into Mexico and will maximize the utility of the ngVLA for Galactic center studies.}
\label{fig:config}
\end{figure}

The ngVLA configuration will consist of $\sim$300 $\times$ 18m antennas\footnote{The number of antennas and configuration shown in Figure~\ref{fig:config} represent the most up-to-date plan for the ngVLA configuration available at time of writing.  For more current information on the ngVLA reference design, we refer readers to \protect{\href{http://ngvla.nrao.edu}{http://ngvla.nrao.edu}}.} with baselines out to at least 300~km ($< 1000$~km) as shown in Figure~\ref{fig:config}. For comparison, the VLA and ALMA have maximum baselines of 36.4 and 16.2~km, respectively.  Given its unique combination of frequency range, angular resolution, and sensitivity, the ngVLA will serve as a transformational new tool in our understanding of how radio jets affect their surroundings.  By combining broadband continuum data with measurements of the cold gas content and kinematics, the ngVLA will quantify the energetic impact of radio jets hosted by gas-rich galaxies as the jets interact with the star-forming gas reservoirs of their hosts.  The ngVLA will also facilitate deep surveys necessary for studying both powerful AGNs at high redshifts and lower-power AGNs associated with low-mass galaxies or MBHs with $M_{\mathrm{BH}} < 10^6$ M$_{\odot}$ in the nearby universe.  These are two key observational frontiers in AGN science that hold important clues for our understanding of MBH-galaxy co-evolution, yet present extreme observational challenges for current instruments.

In Section~\ref{sec:science} of this study, we summarize important AGN science questions on topics including jet-ISM feedback, AGNs hosted by low-mass galaxies, and the role of AGNs in shaping galaxy evolution. In Section~\ref{sec:simulations}, we present simulated ngVLA continuum observations, and discuss the impact of the proposed ngVLA design on future AGN feedback and galaxy evolution studies. Computational considerations are presented in Section~\ref{sec:algorithms}. We examine the synergy between the ngVLA and a variety of existing and future telescopes over a wide range of wavelengths and observational regimes in Section~\ref{sec:synergy}. In Section~\ref{sec:summary}, we summarize the main results of our study and provide suggestions for future work. 

\section{AGN Science in the ngVLA Era}
\label{sec:science}

\subsection{Jet-ISM Feedback}
\label{sec:science_jet_ism_fback}
We still lack a fundamental understanding of exactly how radio jets transfer energy to their surroundings, how much energy is transferred to the different gas phases, and under what conditions significant positive/negative feedback is produced.  We know that radio jets may deposit energy into their surroundings through a number of mechanisms including heating, shocks, and/or turbulence \citep{fabian+12, alatalo+15, soker+16}, and may also directly couple to gas in their surroundings and physically expel it (e.g., \citealt{morganti+13}).  However, the details of these processes -- and under which conditions and environments different mechanisms dominate -- remain unknown.  This is primarily due to the observational challenges of identifying systems with jet-ISM feedback (e.g., sensitivity, angular resolution, and the need for extensive multi-wavelength data) that have prevented large, statistically-complete studies.

\begin{figure*}
\centering
\includegraphics[clip=true, trim=0.25cm 7.15cm 0.15cm 7cm, height=0.365\textwidth]{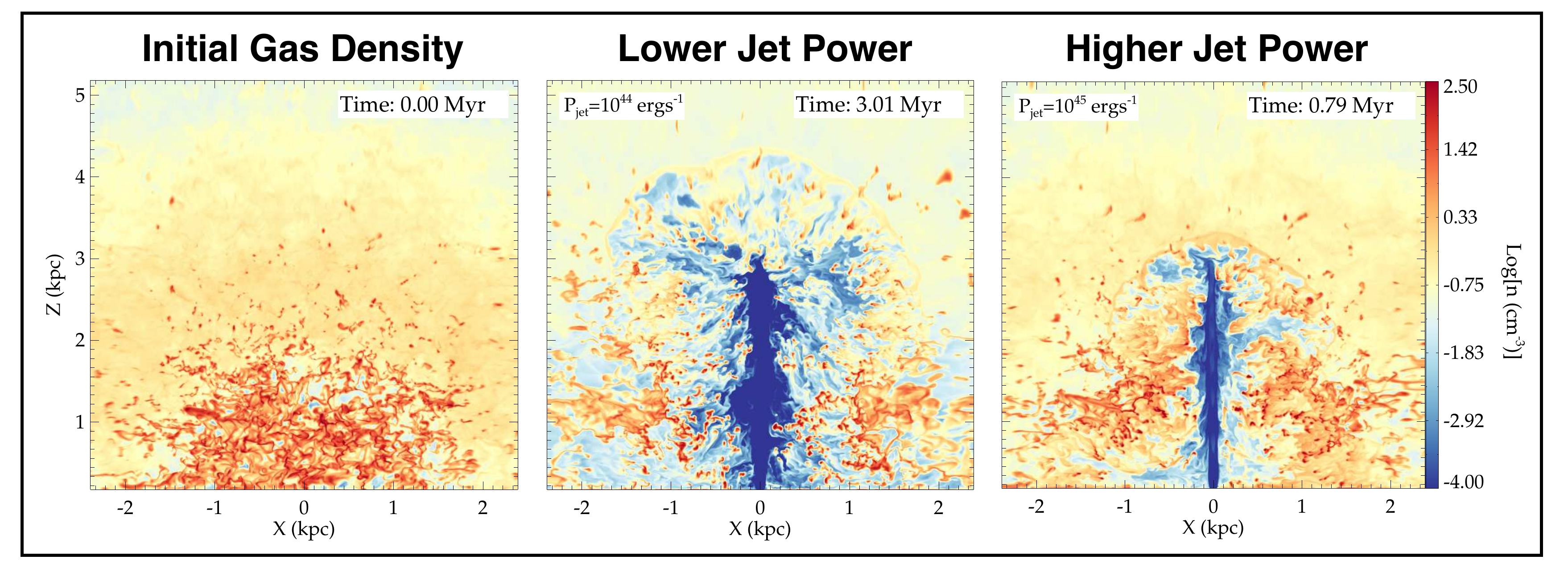}
\caption{Snapshots from relativistic hydrodynamic radio jet simulations \citep{mukherjee+16,mukherjee+17} showing the impact on an identical initial ISM (left) by a radio jet with $P_{\mathrm{jet}} = 10^{44}$~erg~s$^{-1}$ (center) and $P_{\mathrm{jet}} = 10^{45}$~erg~s$^{-1}$ (right). The more powerful radio jet is able to more quickly ``drill'' through the ISM of its host galaxy, while the weaker radio jet is trapped by the ISM and disrupts the surrounding gas for a longer time period and over a larger volume. \\}
\label{fig:jet_sim}
\end{figure*}

Recent observational and theoretical evidence has challenged long-held beliefs that only the most powerful radio AGNs residing in massive elliptical galaxies or at the centers of galaxy clusters are capable of generating significant feedback (e.g., \citealt{alatalo+11, alatalo+15, davis+12, nyland+13, mukherjee+16, mukherjee+17, godfrey+16, querejeta+16, zschaechner+16}), arguing that lower-power radio AGNs may be able to significantly affect the ISM conditions of their hosts through sub-galactic-scale radio jets.  Recent relativistic hydrodynamic simulations of radio jets propagating in a dense ISM (\citealt{mukherjee+16, mukherjee+17, mukherjee+18}; Figure~\ref{fig:jet_sim}) provide strong support for this possibility, demonstrating that while powerful radio jets are able to rapidly ``drill'' through the ISM, lower-power jets become entrained in the ISM and are ultimately able to transfer energy to the ambient gas over a much larger volume and for a longer period of time.  

Radio jet feedback is typically assumed to be {\it negative} in nature in that it involves the destruction, disruption, and/or removal of gas that might otherwise be engaged in current or future star formation.  However, {\it positive} radio AGN feedback -- in which radio jets actually trigger the onset of a burst of star formation or lead to an increased star formation efficiency -- may also occur.  Simulations have shown that the relative importance of negative and positive jet-driven feedback depends on the size distribution and density of the ISM clouds \citep{fragile+04,gaibler+12,gardner+17}, but observational examples of positive radio jet feedback in action are still rare (e.g., \citealt{zinn+13, salome+15, cresci+15, lacy+17, molnar+17}).  

As demonstrated in \citet{lacy+17}, the combination of VLA continuum and ALMA molecular gas imaging helps confirm the presence of positive jet feedback in candidates such as Minkowski's Object.  While future, deep VLA and ALMA observations of radio jets directly interacting with molecular gas in their hosts may help increase the number of known candidates, sensitivity, particularly to the molecular gas emission, will limit the scope of such studies.  A significant advancement in our understanding of the relative roles of positive vs.\ negative AGN feedback in shaping galaxy properties as a function of redshift will therefore require the enhanced sensitivity of the ngVLA.  

\subsection{ISM Content and Conditions}
\label{sec:ISM_content_conditions}
The ngVLA will complement source morphologies and energetics constraints from deep, high-resolution continuum observations with spectral line data that encode information on the ISM content and conditions.  The combination of broadband continuum and spectral line imaging will allow the ngVLA to uniquely probe the energetic impact of radio jets on the ambient cold gas.  Spectral line measurements of molecular and atomic gas on comparable angular scales can be used to identify AGN-driven outflows (as well as gas inflow associated with fueling), perform detailed kinematic studies to gauge the amount of energy injected into the gas via feedback, and address the future evolutionary impact on local/global scales caused by AGN feedback.  These continuum + cold gas ngVLA studies would -- for the first time -- provide us with constraints on the prevalence and energetic importance of jet-ISM feedback in the population of low-luminosity AGNs residing in lower-mass, gas-rich host galaxies.  

\subsubsection{Atomic Gas}
The absorption of atomic hydrogen at 21~cm against background continuum emission associated with a radio AGN provides a powerful means of directly identifying jet-driven outflows and quantifying their effect on the cold ISM (e.g., \citealt{morganti+13}).  \hi\ absorption offers a key advantage over studies of the \hi\ line in emission in terms of detectability, since the detection of \hi\ absorption is independent of redshift and depends solely on the underlying strength of the background continuum source.  In addition, the relatively low spin temperature of \hi\ of $\sim$100 to 150~K \citep{condon+16} makes the detection of emission at high angular resolution difficult or impossible due to brightness temperature sensitivity limitations (see Section~\ref{sec:caveats}).  \hi\ absorption observations, on the other hand, depend only on the brightness temperature of the background continuum source, and may therefore be performed on much smaller (e.g., milliarcsecond) scales.  

\hi\ can only be detected in absorption if it lies in front of the radio plasma, offering important information on the geometry of gas features. In the context of jet-driven feedback, the detection of a blue-shifted spectral component in \hi\ absorption is a robust signature of an outflow, which can be unambiguously distinguished from other possibilities, such as inflow or rotation. Additionally, kinematic constraints from \hi\ absorption observations probe the gas conditions by providing direct measurements of the kinetic energy of any outflow components (\citealt{nyland+13}) as well as characterizing the degree of turbulence \citep{lacy+17}.  

The ability of the ngVLA to observe the \hi\ line will ultimately depend on the lower frequency cutoff of its observing range.  Assuming the ngVLA will observe down to 1.2~GHz, \hi\ studies would be limited to nearby galaxies ($0 < z < 0.1$).  \hi\ absorption surveys of bright (1.4~GHz fluxes $\gtrsim$ a few tens of mJy), nearby radio AGNs with existing radio telescopes have reported detection rates of $\sim$30\% \citep{gereb+14, maccagni+17}, suggesting that blind surveys of \hi\ absorption in even lower-luminosity systems may be possible.  

The possibility of extending the ngVLA's frequency range below 1~GHz (e.g., \citealt{taylor+17}) would greatly expand the redshift range over which \hi\ would be observable with the ngVLA.  We refer readers to \citet{morganti+15a} for a more detailed discussion of the prospects for \hi\ absorption studies with future radio telescopes being designed to operate over more favorable frequency ranges for \hi\ science, such as the Square Kilometre Array (SKA) and its pathfinders.  

\subsubsection{Molecular Gas}
The identification of jet-driven molecular outflows provides an important tool for improving our understanding of the multiphase nature of AGN feedback (e.g., \citealt{rupke+13, emonts+14, sakamoto+14, alatalo+15,morganti+15b}).  Molecular outflows may be identified on the basis of their spectral line shapes, such as the presence of broad wings or a shifted component (e.g., NGC\,1266; \citealt{alatalo+11}), or a P Cygni profile \citep{sakamoto+09}.  A survey of the cold gas properties of a large statistical sample of AGNs spanning a wide range of environments, host galaxy morphologies, MBH masses, and nuclear activity classifications (e.g., high/low Eddington ratios, jetted/radio quiet, Compton thick/unobscured, etc.) would ultimately help establish an observationally-motivated model for the cosmic importance of outflows launched by active nuclei.  While the focus of this work is on ngVLA studies of jet-driven AGN feedback, we note that the ngVLA will also identify outflows driven by radiative winds launched by AGNs operating in the quasar mode.  

In the absence of a detection of an outflow component, a significant increase in the turbulence of the gas, or a substantial change in star formation efficiency/depletion time in the vicinity of the AGN (e.g., at the location the jet; \citealt{lacy+17}), may also provide an indirect means of studying more subtle feedback effects (e.g., \citealt{alatalo+15,oosterloo+17}).  As mentioned in Section~\ref{sec:science_jet_ism_fback}, molecular gas and continuum estimates of the energetics of the outflow and jet, as well as temperature/density estimates in systems where an adequate range of molecular species/transitions have been observed to allow the gas chemistry to be modeled, can be directly compared with state-of-the art simulations, such as those shown in Figure~\ref{fig:jet_sim}, to more deeply explore the feedback physics.

The lowest energy transitions of the CO molecule, CO(1--0), CO(2--1), and CO(3--2) at rest frequencies of 115.2712, 230.5380, and 345.7960~GHz, respectively, trace the total molecular gas reservoir at relatively low densities ($n_{\mathrm{H}_2} \sim 10^3$~cm$^{-3}$).  The CO(1--0) line will be accessible to the ngVLA over the redshift ranges $0< z\lesssim 0.5$ and $z\gtrsim 1.5$.  The gap from $z \approx 0.5$ to $1.5$ is due to the high telluric opacity of molecular oxygen that precludes ground-based observations from 52 to 68 GHz.  Observations of the CO(2--1) line will be possible from $1 \lesssim z\lesssim 2$ and $z\gtrsim 3.5$, and the CO(3--2) line will be accessible over the range $2 \lesssim z\lesssim 4$ and also at $z \gtrsim 6$ (though see Section~\ref{sec:caveats} regarding important caveats).  We note that none of the low-$J$ CO lines will be observable from $z = 0.5 - 1.0$, though transitions of other species probing denser gas, such as SiO and CS, will be accessible.  For a graphical description of the redshift dependence of a wide variety of molecular gas species and transitions observable with the ngVLA, we refer readers to Figures 2 and 9 in \citet{casey+15}.

\subsubsection{Important Caveats}
\label{sec:caveats}
The long ngVLA baselines on scales of a few hundred km -- although advantageous for pinpointing AGNs and resolving their continuum morphologies -- will naturally lead to poor brightness temperature sensitivity, making it virtually impossible to study CO at the maximum angular resolution afforded by the array. With maximum ngVLA baselines of $\sim$500~km in the north-south direction (see Figure~\ref{fig:config}), the maximum angular resolution\footnote{The angular resolution of an interferometer may be estimated by $\theta_{\mathrm{FWHM}} \approx \lambda/B_{\mathrm{max}}$ radians, where $\theta_{\mathrm{FWHM}}$ is the full-width at half maximum of the synthesized beam.  $\lambda$ and $B_{\mathrm{max}}$ are the observing wavelength and maximum baseline, respectively, and are both measured in meters.} at 93~GHz will be $\sim$2~mas.  Even when tapered down to a resolution of $\sim$10~mas, observations in this band will have a brightness temperature sensitivity\footnote{The brightness temperature sensitivity of a point source is defined as $\sigma_{T_\mathrm{B}} = \left( \frac{S}{\Omega_{\mathrm{A}}} \right) \frac{\lambda^2}{2k}$, where $S$ is the flux density in units of W~m$^{-2}$~Hz$^{-1}$, $k$ is the Boltzmann constant ($1.38 \times 10^{-23}$~Jy~K$^{-1}$), and $\lambda$ is the observing wavelength in meters.  The quantity $\Omega_{\mathrm{A}}$ is the beam solid angle and defined as $\Omega_{\mathrm{A}} = \frac{\pi \theta^2_{\mathrm{FWHM}}}{4 \ln(2)}$, where $\theta_{\mathrm{FWHM}}$ is the angular resolution in units of radians.} of $\sigma_{T_\mathrm{B}} \sim 350$~K for an integration time of 1~hr, a channel width of 10~km~s$^{-1}$, and robust weighting \citep{Selina+17}.  The minimum excitation temperature of the CO(1--0) line is 5.53~K, far below the brightness temperature sensitivity of the ngVLA in its highest frequency band (70 to 116~GHz; \citealt{Selina+17}).  Thus, significant tapering of the data, as well as the application of new weighting schemes, will be necessary for the detection of the low-$J$ CO transitions with the ngVLA.  Given the importance of the detectability of these lines to jet-ISM feedback studies and the key science goals defined in \citet{bolatto+17}, a significant investment in the development of new weighting algorithms is clearly warranted.  

At high redshifts, the increasing influence of the cosmic microwave background (CMB) negatively impacts the detectability of the CO molecule, particularly the lower-$J$ CO transitions.  The increasing CMB temperature at high redshifts\footnote{The redshift dependence of the CMB temperature follows the relation $T_{\mathrm{CMB}} = T^{z=0}_{\mathrm{CMB}} \times (1 + z)$, where $T^{z=0}_{\mathrm{CMB}} = 2.73$~K is the CMB temperature at $z = 0$.} both reduces the contrast of the CO emission against the background (particularly in cold molecular clouds with $T_{\mathrm{kinetic}} \sim 20$~K) and changes the shape of the CO spectral line energy distribution by exciting a greater proportion of higher-$J$ rotational levels  (e.g., \citealt{da_cunha+13, zhang+16}).  \citet{da_cunha+13} concluded that this will become particularly problematic for observations of cold molecular clouds in non-starbursting, high-redshift galaxies at $z \sim 4$, where $T_{\mathrm{CMB}} = 13.65$~K.  We point out that for non-starbursting (i.e. main sequence) galaxies a non-negligible evolution of the dust-temperature has been reported (e.g., \citealt{magnelli+14, magdis+12, bethermin+15}), implying that the most pessimistic scenarios in \citet{da_cunha+13} may not apply. However, significantly lower ISM metallicities in high-$z$ galaxies will be an additional factor that leads to suppressed CO emission.

In molecular clouds in the vicinity of AGNs and jets, CMB heating is expected to be less problematic since AGN heating may increase the gas kinetic temperature to hundreds of degrees \citep{matsushita+98, krips+08, viti+14, glenn+15, richings+18}.  Detailed studies of the cold gas properties in the vicinity of AGNs with ALMA prior to the construction of the ngVLA will be essential for quantifying the impact of CMB heating on observing cold gas in the nuclei of distant galaxies.      

\begin{figure*}
\centering
\includegraphics[clip=true, trim=0.1cm 5cm 2.2cm 5.9cm, height=0.475\textwidth]{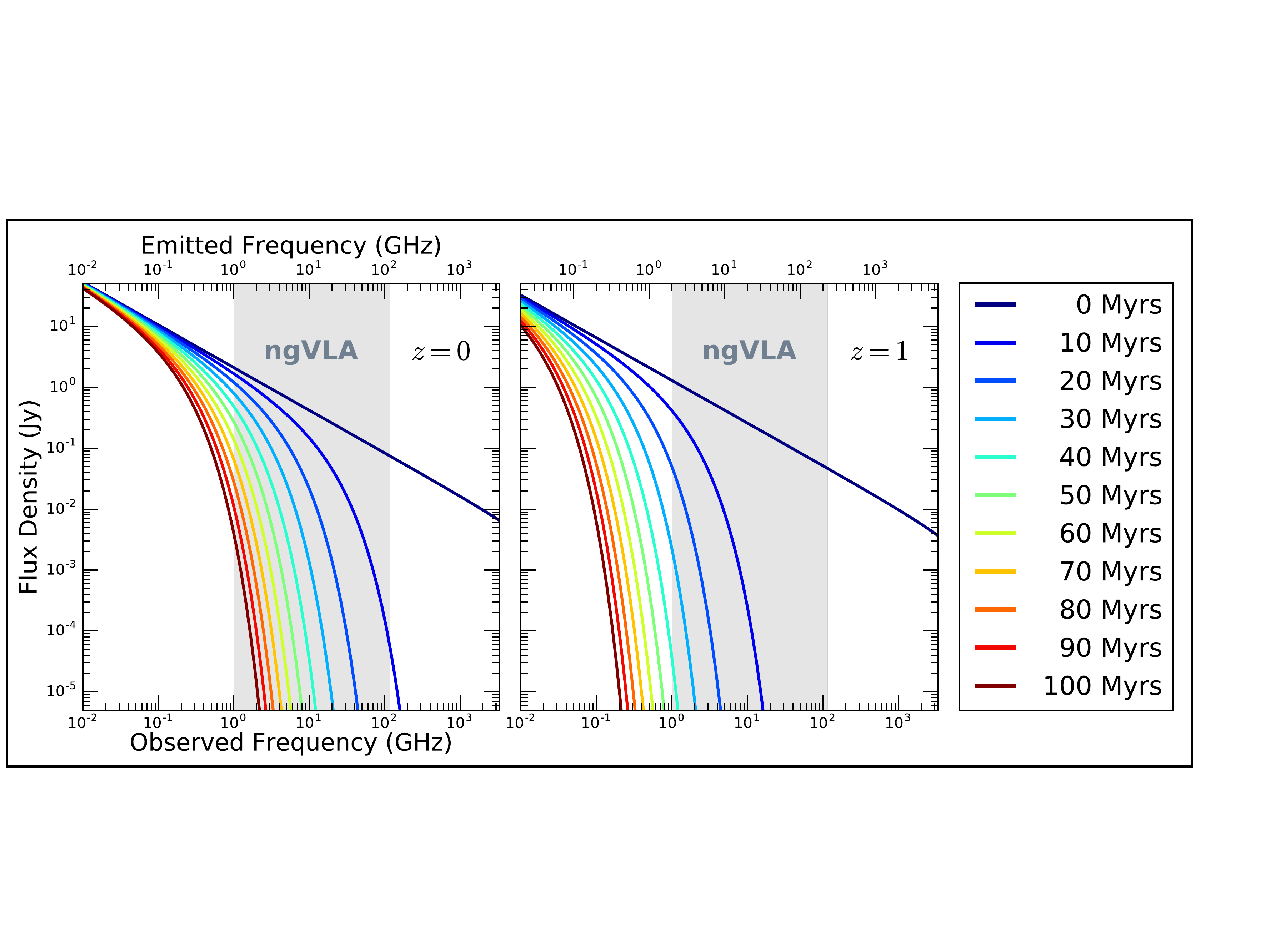}
\caption{Example of JP model \citep{jaffe+73} spectral ages calculated using the BRATS software \citep{harwood+13} demonstrating the need for ngVLA observations spanning a wide range of frequencies. The left and center panels correspond to redshifts of 0 and 1, respectively.  Standard spectral age model parameters were used (injection index $=0.7$, magnetic field strength of 1~nT, and electron minimum and maximum Lorentz factors of 10 and 10$^6$, respectively).  The flux density values shown on the $y$-axis have been arbitrarily scaled.  Because of its advantages of wide frequency range and angular resolution compared to the SKA, the ngVLA will uniquely excel in studies of low-redshift radio AGNs that are young, or higher-redshift AGNs that are embedded in dense environments.\\}
\label{fig:JP}
\end{figure*}

\subsection{Radio Spectral Ages}
\label{sec:ages}
In theory, for an electron population in a fixed magnetic field with an initial energy distribution described by a power law such that $N(E) = N_0 E^{-\delta}$, the energy losses scale as $\tau = \frac{E}{dE/dt} \propto 1/E \propto 1/\nu^{2}$, ultimately leading to a preferential cooling of higher energy electrons.  In the absence of any further particle acceleration, this produces a spectrum that becomes increasingly curved over time allowing us to determine the characteristic age of a source (e.g., \citealt{myers+85, harwood+13}).

In Figure~\ref{fig:JP}, we show an example of JP model \citep{jaffe+73} spectral ages calculated using the BRATS software\footnote{\protect{\href{http://www.askanastronomer.co.uk/brats/}{http://www.askanastronomer.co.uk/brats/}}} \citep{harwood+13} for an arbitrary jetted radio AGN at $z = 0$ and $z = 1$.  As indicated by the spectral age curves shown in this figure, the ngVLA will excel in studies of radio AGNs spanning a wide range of ages at low redshift, as well as radio AGNs that are young or embedded in dense environments at higher redshifts.  We emphasize that, unlike other next-generation radio facilities that will operate primarily at lower frequencies (Section~\ref{sec:synergy_SKA}), the planned observing frequency range of the ngVLA from 1 to 116~GHz makes it uniquely equipped for measuring the ages of {\it young} radio AGNs potentially engaged in jet-ISM feedback.  As shown by results from the Australia Telescope 20 GHz (AT20G) survey \citep{murphy+10}, measurements in this frequency range are needed to adequately model radio spectral energy distributions \citep{sadler+06,sadler+08}. This is particularly important for modeling the ages of young, low-redshift sources less than 10~Myrs old.  Lower frequency radio continuum data in the MHz range are important for constraining the ages of high-$z$ sources; however, the inclusion of the lowest-frequency ngVLA bands down to $\sim$1~GHz would provide sufficient frequency coverage for measuring source ages as old as 30 to 40~Myrs at $z \sim 1$. 

\subsection{Young Radio AGNs}
Young, compact radio sources associated with accreting MBHs  represent a key phase in the life cycles of jetted AGNs.  These objects also hold important clues about radio AGN triggering and duty cycles, which are still poorly understood \citep{tadhunter+16}. These sources include Gigahertz-peaked Spectrum (GPS) sources and High-frequency Peakers (HFP), which have convex radio spectral energy distributions (SEDs) that peak from 1 to a few GHz and have jet extents ranging from a few pc to tens of kpc, respectively  \citep{odea+98, orienti+16}.  The wide frequency range and high angular resolution of the ngVLA is well tuned for studies of young radio AGNs that have recently (re)ignited their central engines within the past 10$^2$ to 10$^4$ years, particularly those with sub-galactic jet extents of $\sim$10~pc to a few kpc (see Figure~\ref{fig:jet_size_z} and Section~\ref{sec:obs_imp}).

An open question regarding young radio AGNs is their energetic impact on the ISM of the host galaxy.  A growing number of studies have reported detections of multiphase outflow signatures associated with young radio AGNs engaged in jet-ISM feedback (e.g., \citealt{chandola+11, holt+11, morganti+13}), but the importance of this feedback in the context of galaxy evolution remains unclear. The ngVLA  will address this issue through both continuum and spectral line observations.  Broad-band continuum measurements of the turnover frequency of the radio SEDs of young radio sources can give a direct handle on the density distribution of the ISM for sources at low redshift \citep{bicknell+97, bicknell+18, jeyakumar+16}.  Follow-up spectral line observations probing the conditions of the ISM in the vicinity of the jets of the young radio AGNs will provide further constraints on the kinematics of the gas, thus probing the energetic impact of the jet-driven feedback.

\begin{figure*}
\centering
\includegraphics[clip=true, trim=0.05cm 5.5cm 0.05cm 7cm, height=0.36\textwidth]{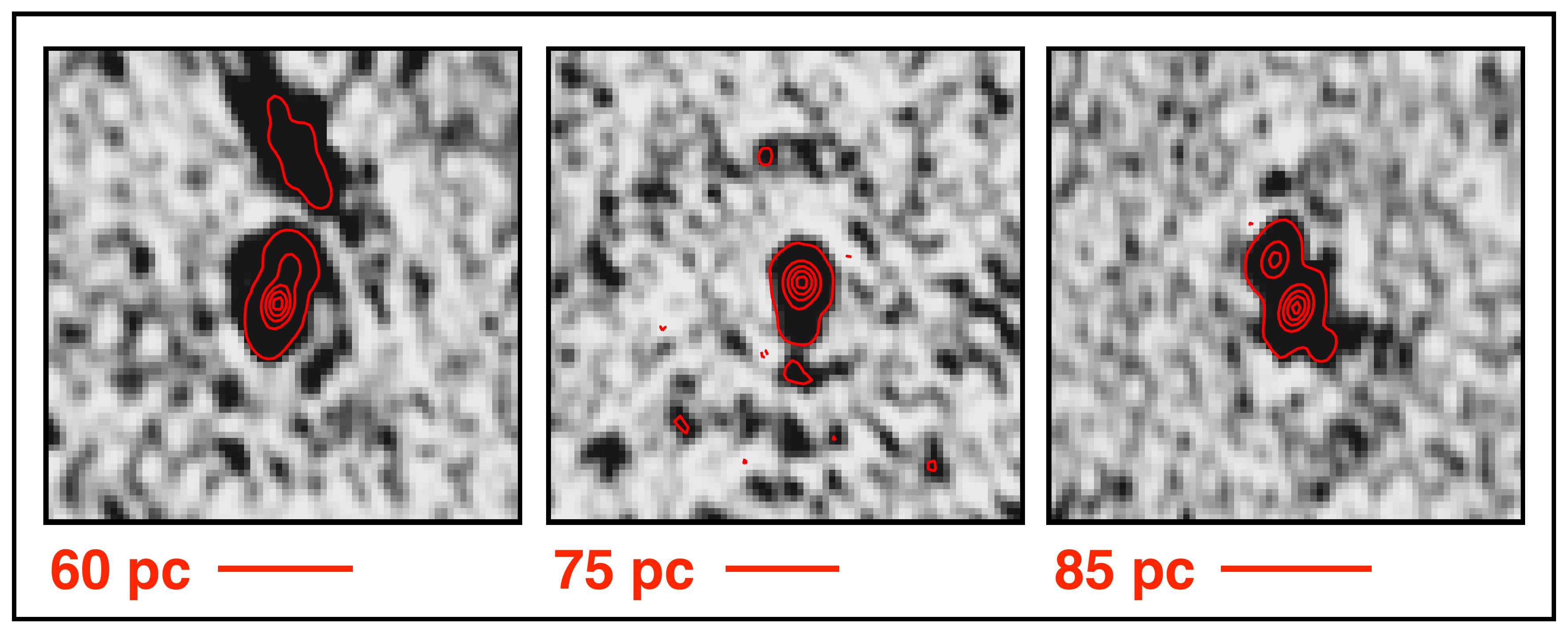}
\caption{Example VLBA continuum image cutouts (each with dimensions of 40 $\times$ 40 mas) from Patil et al.~(in preparation) of young, compact radio AGNs with source extents of tens of pc.  These sources have redshifts in the range of $0.5 \lesssim z \lesssim 2$ and are drawn from the sample presented in \citet{lonsdale+15}.  The Lonsdale sample selects heavily obscured, powerful AGNs embedded in luminous starbursting galaxies at intermediate redshifts with compact radio emission based on their properties from the Wide-Field Infrared Survey Explorer \citep{wright+10} and the NRAO--VLA Sky Survey \citep{condon+98}.  The high collecting area and angular resolution of the ngVLA will facilitate deep, spatially-resolved surveys of compact radio jets with comparable extents in a wide range of systems, providing new insights into the energetic impact of young radio jets under a wide range of host galaxy conditions and properties. \\}
\label{fig:pallavis_sources}
\end{figure*}

In Figure~\ref{fig:pallavis_sources}, we show examples of candidate young radio AGNs with jet extents of 60 to 85~pc based on high-resolution continuum observations with the Very Long Baseline Array (Patil et al., in preparation). These radio AGNs were drawn from the sample of extremely infrared luminous sources identified in \citet{lonsdale+15} that are believed to be in a unique evolutionary stage just after the (re)ignition of the radio AGN but while the host galaxy is still experiencing substantial starburst activity.  The ngVLA offers an ideal combination of angular resolution, frequency range, and sensitivity to efficiently resolve the structures of young radio AGNs, such as the sources highlighted in Figure~\ref{fig:pallavis_sources}, and constrain their ages through broadband continuum observations.

\subsection{Low-mass Galaxies}
\label{sec:dwarf_galaxies}
The local volume of nearby galaxies includes 869 identified galaxies with individual distance estimates within 11~Mpc, about 75\% of which are dwarf galaxies \citep{karachentsev+13}.  Determining the occupation fraction of central MBHs in this population of nearby dwarf ($M_{*} \lesssim 10^9$~M$_{\odot}$) and low-mass ($10^9 \lesssim  M_{*} \lesssim 10^{10}$~M$_{\odot}$) galaxies is inherently difficult due to their expected low masses and weak accretion signatures.  However, progress in this area would have a profound impact on our understanding of the formation of MBH seeds at high redshift and the connection between the cosmic assembly and evolution of MBHs and galaxies.  Recent studies of the nearby AGN population (e.g., \citealt{mezcua+16}) have provided indirect evidence that accreting central MBHs with masses of $M_{\mathrm{BH}} \lesssim 10^5$~M$_{\odot}$ may be common among low-mass galaxies at low redshifts, thus motivating future searches for their faint synchrotron signatures through deep radio continuum observations.

Another population of galaxies for which the presence of MBHs remains largely unexplored is that of low-surface-brightness (LSB) galaxies, which are characterized by diffuse stellar disks and high dark matter fractions \citep{impey+97}.  Studies of AGNs hosted by the subset of bulge-dominated LSB galaxies have reported estimated MBH masses in the range of $10^5 \lesssim  M_{*} \lesssim 10^7$~M$_{\odot}$ that tend to fall below the standard relationship between MBH mass and stellar spheroid mass \citep{ramya+11, subramanian+16}.  Dwarf galaxies may also lie offset from the $M_{\mathrm{BH}}-M_{\mathrm{sph,*}}$ relation (e.g., \citealt{graham+15, nguyen+17}).  If verified through larger statistical studies utilizing accurate MBH mass measurements with dynamical constraints, this might indicate a difference in the connection between MBH and galaxy formation pathways for different host galaxy populations, a prospect that has important implications for our understanding of cosmic assembly.

In addition to the importance of constraining the MBH occupation fraction in lower-mass galaxies to gain insights into MBH seed formation and growth, this MBH population may also energetically impact the properties of their hosts.  There is strong empirical -- albeit circumstantial -- evidence that AGN feedback may play an important evolutionary role in low-mass galaxies at high redshift and perhaps also their low-redshift counterparts, nearby dwarf galaxies \citep{silk+17}.  Observational studies identifying candidate AGNs hosted by low-mass/dwarf galaxies have become increasingly common in the literature (\citealt{barth+08, greene+04, greene+07b, reines+13, moran+14, lemons+15, sartori+15, nucita+17}).  Recent analytical models exploring prospects for AGN feedback in dwarf galaxies suggest that it may provide an efficient mechanism for the displacement of gas from the host galaxy \citep{dashyan+18}. This scenario may be most plausible in galaxies with stellar masses in the range of $10^7 \lesssim  M_{*} \lesssim 10^9$~M$_{\odot}$ when preceded by substantial supernova feedback capable of rarefying the ISM, thus making it more susceptible to disruption via AGN feedback \citep{prieto+17, hartwig+18}.  Observational evidence for AGN-driven feedback in low-mass and dwarf galaxies is rare, though recent spatially-resolved spectroscopic studies offer tentative support (e.g., \citealt{penny+18}).  Radio continuum identifications of jetted AGNs hosted by dwarf galaxies with the potential to impart feedback on their hosts is also challenging with currently available instruments (e.g, \citealt{nyland+16, padovani+16}), which lack adequate collecting area and angular resolution. 

Only a handful of candidate jetted AGNs hosted by low-mass or dwarf galaxies are known, such as Henize 2-10 \citep{reines+11} and NGC\,404 \citep{binder+11, seth+10, nyland+12, paragi+14, nguyen+17}.  In a recent study, \citet{nyland+17} presented multiwavelength evidence of shocks associated with a confined jet spanning 17~pc in the center of the nearby dwarf galaxy NGC\,404, raising the possibility of extremely weak, localized feedback from the AGN.  However, to our knowledge, a convincing observational example of jet-ISM feedback in action in a low-mass or dwarf galaxy has yet to be identified given the capabilities of current instruments.  The ngVLA, with its order of magnitude increase in sensitivity and angular resolution compared to the VLA, will greatly improve our ability to pinpoint and study the faint accretion signatures of MBHs with masses comparable to those of MBH seeds in the nuclei of low-mass, dwarf, and LSB galaxies and constrain their energetic imprint on their hosts.  To further explore the advancements in this area that will be made possible by the ngVLA, we present simulations of a $\sim 3\times$ more distant ($D=11$~Mpc) analog to the confined jet with an extent of $\sim$10~pc in the center of NGC\,404 in Section~\ref{sec:NGC404_sims} and Figure~\ref{fig:NGC404_sim}.

\subsection{Polarization}
\label{sec:pol}
Full-polarization continuum imaging with the ngVLA will provide additional key insights into radio AGN physics and the coupling between jets and their surroundings. Of particular interest are measurements of the degree of Faraday rotation associated with AGNs that will probe the magnetic field orientation and the interaction between radio jets and the surrounding magneto-ionic medium (e.g., \citealt{gomez+11, osullivan+12}).  Faraday rotation refers to the rotation of the electric vector position angle as an electromagnetic wave travels through a magnetized plasma.  Observationally, this effect is quantified by the rotation measure (RM), which depends linearly on $\lambda^2$, and is defined as: 

\begin{equation}
\mathrm{RM} = 812 \int n_{\mathrm{e}}\, B_{\mathrm{LOS}}\, dl \,\,\,\, [\mathrm{rad}\,\,\mathrm{m}^{-2}],
\end{equation}

\noindent where $n_{\mathrm{e}}$ is the electron number density (cm$^{-3}$), $B_{\mathrm{LOS}}$ is the line-of-sight component of the magnetic field (milli-Gauss) and $dl$ is the path length (pc).  
The physical origin of the Faraday rotation effect observed in a subset of the radio AGN population is still a subject of debate.  Possible explanations include internal Faraday rotation caused by the radio jet itself, the presence of helical magnetic fields near the launch point of the jet, external depolarization caused by a sheath of magnetized plasma surrounding the jet, and the presence of foreground clouds interacting with the jet through feedback (\citealt{gomez+11}, and references therein).  Distinguishing between these effects requires high sensitivity for the detection of inherently faint polarization signatures, high angular resolution for measuring spatial gradients in the RM, broadband observing capabilities from centimeter to millimeter wavelengths for identifying sources with high RMs, and multi-epoch observations capable of providing constraints on temporal variability. 

The combination of high sensitivity and angular resolution, coupled with frequency coverage from 1 to 116~GHz, make the ngVLA an ideal tool for probing magnetic fields, jet formation, and feedback in radio AGNs through their polarimetric properties.  As also highlighted in \citet{casey+15}, the uniqueness of the ngVLA for polarization studies lies in its ability to identify sources with extreme RMs ($\gtrsim 10^4$~rad~m$^{-2}$) at millimeter wavelengths that are typically depolarized in the centimeter-wave regime.  These objects are rare, but examples include Sgr~A$^{*}$ \citep{macquart+06, marrone+07}, M81$^{*}$ \citep{brunthaler+06}, and 3C273 \citep{attridge+05}. \citet{arshakian+11} estimate an optimal observing frequency of $\nu > 30$~GHz ($\lambda < 10$~mm) for the identification and study of sources with extreme RMs, well within the proposed frequency range of the ngVLA.   

\subsection{The Galaxy Evolution Connection}
Despite its well-established importance, the role of energetic feedback driven by AGNs in regulating galaxy formation and evolution remains a poorly understood aspect of galaxy assembly.  The exceptional sensitivity, resolution, and frequency range of the ngVLA will place new constraints on the prevalence of AGN feedback operating in {\it both} the jet and radiative modes out to high redshifts.  These studies will facilitate comprehensive tests of observational predictions from cosmological simulations that aim to model the processes that influence MBH growth across cosmic time (e.g. \citealt{volonteri+16, martin+18}).  In Sections~\ref{subsec:AGNatHighz} and \ref{subsec:quasar_fback} below, we briefly discuss prospects for surveying high-redshift AGNs in deep continuum observations as well as placing tight constraints on the energetic importance of radiative-mode feedback through observations of the Sunyaev Zel'dovich effect in the vicinity of powerful quasars.

\subsubsection{Deep Continuum Surveys}
\label{subsec:AGNatHighz}
Large surveys of radio AGNs out to $z \sim 2$ have become increasingly common and indicate a high space-density of radio AGNs during this epoch (e.g., \citealt{luchsinger+15}).  However, detections of radio AGNs at $z \gtrsim 4$ remain rare (e.g., \citealt{mortlock+11, van_breugel+99}).  Recent studies have hinted at the importance of deep, high-resolution radio continuum observations for identifying radio AGNs at high redshift and determining their influence on early MBH and galaxy growth and evolution (e.g., \citealt{rujopakarn+16, rujopakarn+18, maini+16}).  

The high sensitivity and angular resolution of the ngVLA will capture AGN and SF emission in an extinction-free manner at the peak epoch of cosmic assembly.  Surveys with the wide bandwidth of the ngVLA will also facilitate resolved spectral index (Section~\ref{sec:ages}) and Faraday rotation (Section~\ref{sec:ages}) studies encompassing large numbers of sources. A survey with the ngVLA carried out in a similar manner to the VLA Sky Survey (VLASS\footnote{\protect{\href{https://science.nrao.edu/science/surveys/vlass}{https://science.nrao.edu/science/surveys/vlass}}}) would be significantly more efficient due to the wider field of view of the 18m dishes and higher collecting area compared to the VLA. An all-sky survey with the same total observing time as VLASS ($\approx$6000~hours) could be $\approx$10 times deeper, reaching the star forming galaxy population at $z\sim 1$ and allowing spectral index and polarization studies of $\sim$10$^7$ AGN at sub-arcsecond angular resolution.

We emphasize that in conjunction with high-resolution ngVLA studies, particularly at frequencies of tens of GHz and below, low-frequency surveys with the SKA and its pathfinders are certain to make significant contributions to our understanding of the high-$z$ radio AGN population as well.  We discuss synergy between the ngVLA and the SKA in Section~\ref{sec:synergy_SKA}.

\subsubsection{Quasar Feedback Signatures}
\label{subsec:quasar_fback}
Although this paper is primarily focused on ngVLA studies of AGN feedback operating in the radio mode, the ngVLA will also facilitate studies of quasar-mode feedback at high redshift.  Based on outflow models by \citet{faucher+12}, \citet{nims+15} argue that accretion-disc winds from quasars may leave an observational signature in radio synchrotron emission when shocks drive into the ISM. They estimate that the radio luminosity can exceed the far infrared-radio correlation of normal star forming galaxies \citep{condon+92}, even for radio-quiet quasars. Quasar feedback models also predict the presence of hot thermal bubbles on scales of $\sim$10$-$100~kpc in the vicinity of distant quasars.  A detection of the signature of these bubbles by measuring the Sunyaev Zel'dovich (SZ) decrement in the vicinity of powerful quasars (e.g., \citealt{chatterjee+08, rowe+11, dutta_chowdhury+17}) would provide a direct observational confirmation of quasar mode feedback.  This would have important consequences for our understanding of  thermal winds launched by quasars, and provide constraints on the energetics of quasar-mode feedback for future cosmological simulations.  

The frequency range of the ngVLA is ideal for studies of quasar feedback.  For measurements made at several tens of GHz, the contributions by both synchrotron emission at lower frequencies ($\nu \lesssim 10$~GHz) associated with jets or star formation, and dust at higher frequencies ($\nu \gtrsim 150$~GHz), should minimize contamination of the SZ decrement \citep{platania+02}.  Currently, ALMA and the VLA can only usefully constrain the SZ decrement from the 1 or 2 most luminous quasars known. With the ngVLA, we will be able to extend these studies to less luminous quasars, and investigate the dependencies on quasar luminosities and the presence or absence of winds in the optical (as determined by the absorption line properties of the optical spectra). 

Given the superb sensitivity of the ngVLA, it should also be possible to detect both radio jet and quasar mode feedback in action since quasars are often found in groups and clusters (e.g., \citealt{wold+01}), which produce their own SZ signal on larger scales than expected from feedback winds. Any bubbles of relativistic electrons from radio jets will show up as a weakening of the SZ signal in those regions, and indeed a detailed ngVLA study could even help determine the composition of the radio jets \citep{pfrommer+05}.

\section{Simulations}
\label{sec:simulations}
The breadth of AGN science that can be accomplished with the ngVLA will ultimately depend on its sensitivity, frequency range, and the range of spatial scales over which structures may be imaged.  In order to quantify the importance of angular resolution for future studies of jetted AGNs, we have produced simulated ngVLA datasets based on real radio interferometric images of AGNs with extended continuum morphologies.  The simulations are noise-free, and are thus intended to illustrate the performance of the antenna configuration considered in this study and its ability to sample the range of spatial scales present in the input source models.

\subsection{Method and Assumptions}
We produced input model images by masking the original source images with zeros down to approximately the 5$\sigma$ level.  The simulations were performed using the task {\sc SIMOBSERVE} in the Common Astronomy Software Applications package (CASA; \citealt{mcmullin+07}) with the ngVLA antenna configuration shown in Figure~\ref{fig:config}.  To simulate a higher-redshift analog of each radio source, we decreased the image cell size in the input model image by the ratio of the angular diameter distances at the original and new source redshift using the CASA toolkit.  Angular diameter distances were computed using the cosmology sub-package of {\sc Astropy} \citep{astropy+13} with the cosmological parameters from the {\it Planck} satellite \citep{Planck15}.  In addition to simulating observations with the ngVLA, we also performed identical simulations in the VLA A configuration, which has a maximum baseline length of $B_{\mathrm{max}}$ = 36~km.  

\begin{figure}
\centering
\includegraphics[clip=true, trim=3cm 0cm 2cm 0cm, height=0.45\textwidth]{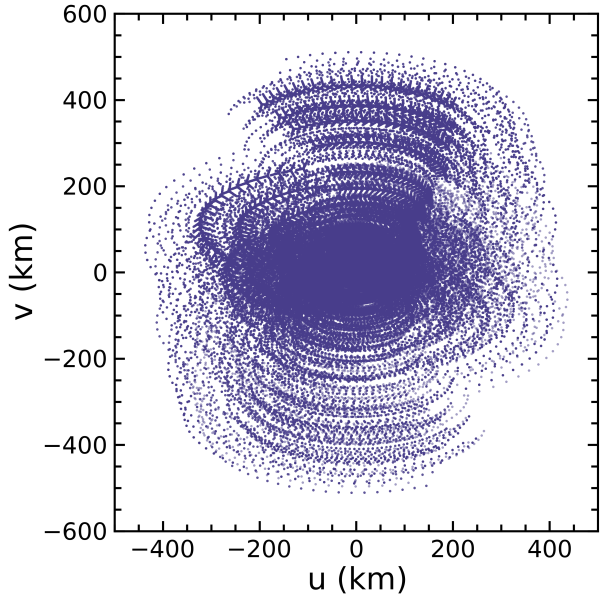}
\caption{Typical $uv$-coverage of a northern declination ($\delta \sim +36^{\circ}$) source from our ngVLA simulations.}
\label{fig:uvcov}
\end{figure}

With maximum east-west baselines of $\sim$300~km and assuming an antenna diameter of 18~meters, a correlator integration time of $t_{\mathrm{int}} < 0.5$~seconds and a channel width of $\Delta \nu_{\mathrm{chan}} < 30$~kHz would be necessary to prevent significant time-averaging loss and bandwidth smearing, respectively, at the half-power point of the primary beam (further technical details are provided in Section~\ref{sec:algorithms}).  
For a ngVLA configuration consisting of 300 antennas, the number of visibilities in a simulated dataset with full frequency and time resolution would be extremely large\footnote{The number of instantaneous visibilities per polarization, channel, and integration time is equal to the number of baselines.  For an array of 300 antennas, the number of baselines is $N_{\mathrm{ant}}(N_{\mathrm{ant}}-1)/2 = 44850$.  For a channel width of $\Delta \nu_{\mathrm{chan}} = 30$~kHz and $t_{\mathrm{int}}=0.5$~sec, an 8-hour simulated dataset with a total bandwidth of only 128~MHz would consist of more than 11 trillion visibilities per polarization.} and demand substantial computing resources for the production of simulations with a realistic spectral and temporal set up.  In order to avoid long computation times in CASA or the need for high-performance computing resources, we performed our simulations with unrealistically coarse spectral and temporal sampling.  Thus, we limited our simulations to a single spectral channel of width 128~MHz and simulated full-synthesis observations spanning a total of 8 hours with a correlator integration time of 20 minutes. 

Images of the simulated datasets were formed using the {\sc CLEAN} task in CASA with the multi-scale, multi-frequency-synthesis algorithm \citep{rau+11} and Briggs weighting with robust = $-1$.  In Figure~\ref{fig:uvcov}, we show the $uv$-coverage of our simulated ngVLA observations for a northern source at an arbitrary location of $\alpha_{\mathrm{J2000}}$~=~01$^{\mathrm{h}}$09$^{\mathrm{m}}$27.000$^{\mathrm{s}}$ and $\delta_{\mathrm{J2000}}$~=~+35$\degr$43$\arcmin$04.70$\arcsec$.  For simplicity, all of the simulated datasets presented in this study were centered at this sky position. We summarize the properties of our simulated sources in Table~\ref{tab:sim_summary}.  
In Figures~\ref{fig:3C28_sim}, \ref{fig:NGC1266_sim}, and \ref{fig:NGC404_sim} we show the results of our VLA and ngVLA continuum simulations of redshifted analogs of three real AGNs representing a wide range of jet spatial scales from tens of pc to hundreds of kpc.

\begin{deluxetable*}{cccCCCCCCCCCCCC}[t!]
\tablecaption{Summary of Input Model Source Properties  \label{tab:sim_summary}}
\tablecolumns{18}
\tablewidth{0pt}
\tablehead{
\colhead{Source} & \colhead{$d_{\mathrm{jet}}$} & \colhead{$\nu$} & \colhead{$z$} & \colhead{$D$} & \colhead{$\theta_{\mathrm{jet}}$} & \colhead{$F_{\mathrm{total}}$} & \colhead{$z^{\prime}$} & \colhead{$D^{\prime}$}  & \colhead{$\theta_{\mathrm{jet}}^{\prime}$} &  \colhead{$F_{\mathrm{total}}^{\prime}$}  \\
\colhead{} & \colhead{kpc} & \colhead{GHz} & \colhead{} & \colhead{(Mpc)} & \colhead{(arcsec)} & \colhead{(mJy)} & \colhead{} & \colhead{(Mpc)} & \colhead{(arcsec)} & \colhead{(mJy)} \\
\colhead{(1)} & \colhead{(2)} & \colhead{(3)} & \colhead{(4)} & \colhead{(5)} & \colhead{(6)} & \colhead{(7)} & \colhead{(8)}  & \colhead{(9)}  & \colhead{(10)}  & \colhead{(11)} 
}
\startdata
3C28 & 150 & 6 & 0.195275 & \nodata & 44.1 & 94.12 & 1.0 & \nodata & 18.22 & $1.25 \times 10^3$\\
NGC\,1266 & 1.5 & 5 & 0.007238 & \nodata & 10.0 & 39.55 & 0.2 & \nodata & 0.44 & 33.67\\
NGC\,404 & 0.017 & 15 & \nodata & 3.1 & 1.1 & 0.29 & \nodata & 10.0 & 0.35 & 27.87\\
\enddata
\tablecomments{Column 1: Source name.  Column 2: The projected linear jet extent, measured from end to end.  Column 3: Original observing frequency of the radio continuum data used to construct the input model for our simulations.  We also center our simulations at this frequency.  Column 4: True source redshift.  All redshifts are given in the 3K background reference frame.  Column 5: Due the close proximity of NGC\,404, determining an accurate cosmological redshift is non-trivial because its observed velocity is dominated by bulk peculiar motions.  Thus, we provide the linear distance for this galaxy instead.  Column 6: Angular jet extent, measurement from end to end, at the redshift or distance given in Column 4 or 5.  Column 7: Total source flux measured at the redshift or distance given in Columns 4 or 5.  Column 8: Simulation redshift.  Column 9: Simulation distance.  Column 10: Angular jet extent, measurement from end to end, at the redshift or distance given in Column 8 or 9.  Column 11: Total source flux measured at the redshift or distance given in Columns 8 or 9.
 }
\end{deluxetable*}

\subsection{Representative Source Models}

\subsubsection{100~kpc-scale Jet}
\label{sec:sims_100kpc_jet}
\begin{figure*}[b!]
\centering
\includegraphics[clip=true, trim=0.05cm 7cm 0.05cm 4cm, height=0.4\textwidth]{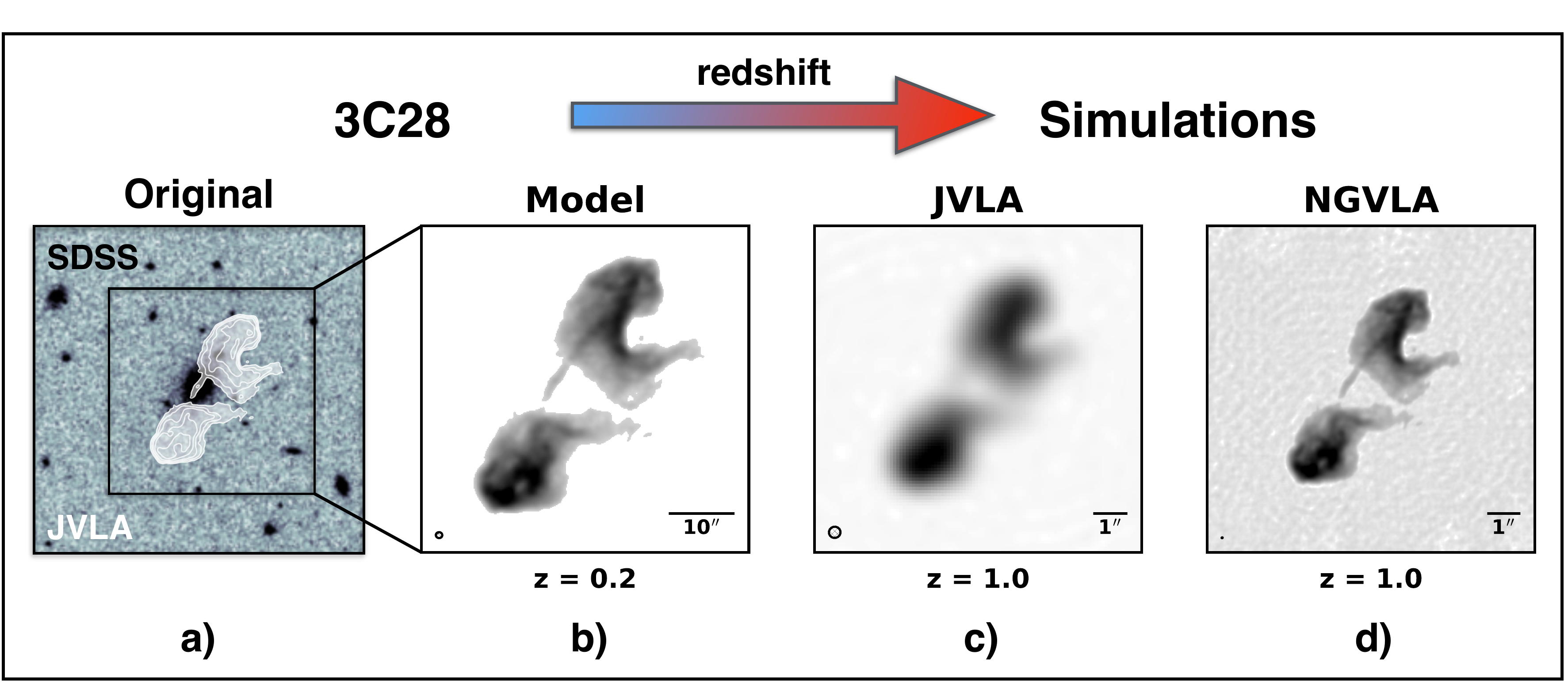}
\caption{Simulated VLA and ngVLA images of a redshifted analog to the radio galaxy 3C28.  {\bf a)} A Sloan Digital Sky Survey DR12 $i$-band optical image is shown in the background colorscale and the multi-configuration (B and C array) VLA $C$-band (6~GHz) radio contours are overlaid in white.  The angular resolution of the radio data is $\theta_{\mathrm{FWHM}} = 1.0^{\prime \prime}$ and the extent of the radio lobes from end to end is 45$^{\prime \prime}$ (150~kpc).  The 3C28 VLA data were originally published in \citet{harwood+15}.  {\bf b)} Model radio image of 3C28 based on the original data shown in panel a) and shown at the true redshift of the source of $z = 0.2$. {\bf c)}  Simulated VLA A-configuration map of a 3C28 analog redshifted to $z = 1.0$ at 6~GHz with $\theta_{\mathrm{FWHM}} = 0.34^{\prime \prime}$.  {\bf d)} Simulated map of a 3C28 analog at $z = 1.0$ as it would appear if imaged with the ngVLA at 6~GHz.  The angular resolution is $\theta_{\mathrm{FWHM}} = 0.028^{\prime \prime}$.
\\}
\label{fig:3C28_sim}
\end{figure*}

The simulations shown in Figure~\ref{fig:3C28_sim} explore prospects for studies of classical radio galaxies with properties similar to 3C28, an example of an AGN with prominent lobes with an extent on the order of 100~kpc that is located at the center of a galaxy cluster at $z \sim 0.2$.  Radio galaxies like 3C28 are commonly classified according to the historical \citep{fanaroff+74} type I (FRI; ``edge darkened'') and II (FRII; ``edge brightened'') categories, with FRI sources typically associated with lower radio powers and inefficient MBH accretion compared to the FRII class. While the radio morphology of 3C28 is somewhat unusual for a radio galaxy, \citet{harwood+15} argue on the basis of their spatially-resolved spectral aging analysis that 3C28 is consistent with an FRII radio galaxy produced by an AGN that shut down $\sim$6-9~Myrs ago.

The complex, extended morphology of 3C28 spans a wide range of spatial scales and required multi-configuration VLA observations.  3C28 thus serves as an excellent test case for simulating the performance of the ngVLA antenna configuration.  As shown in Figure~\ref{fig:3C28_sim}, the ngVLA is able to capture the full range of structures visible in the original image model for an analog of 3C28 redshifted out to $z \sim 1.0$.  VLA observations of the same redshifted input model are less successful at recovering the full range of morphological features, though the double lobed nature of the radio emission is evident nonetheless.  

This simple analysis suggests that, for spatially-resolved studies of the radio spectral index or spectral age modeling, the ngVLA would in principle have a small but important advantage over observations of identical sources with the VLA in the A-configuration.  In Section~\ref{sec:obs_imp}, we evaluate the feasibility of our redshifted 3C28 ngVLA and VLA simulations with an emphasis on the implications of the surface brightness sensitivity of the ngVLA. 

\begin{figure*}[t!]
\centering
\includegraphics[clip=true, trim=0.05cm 7cm 0.05cm 4cm, height=0.4\textwidth]{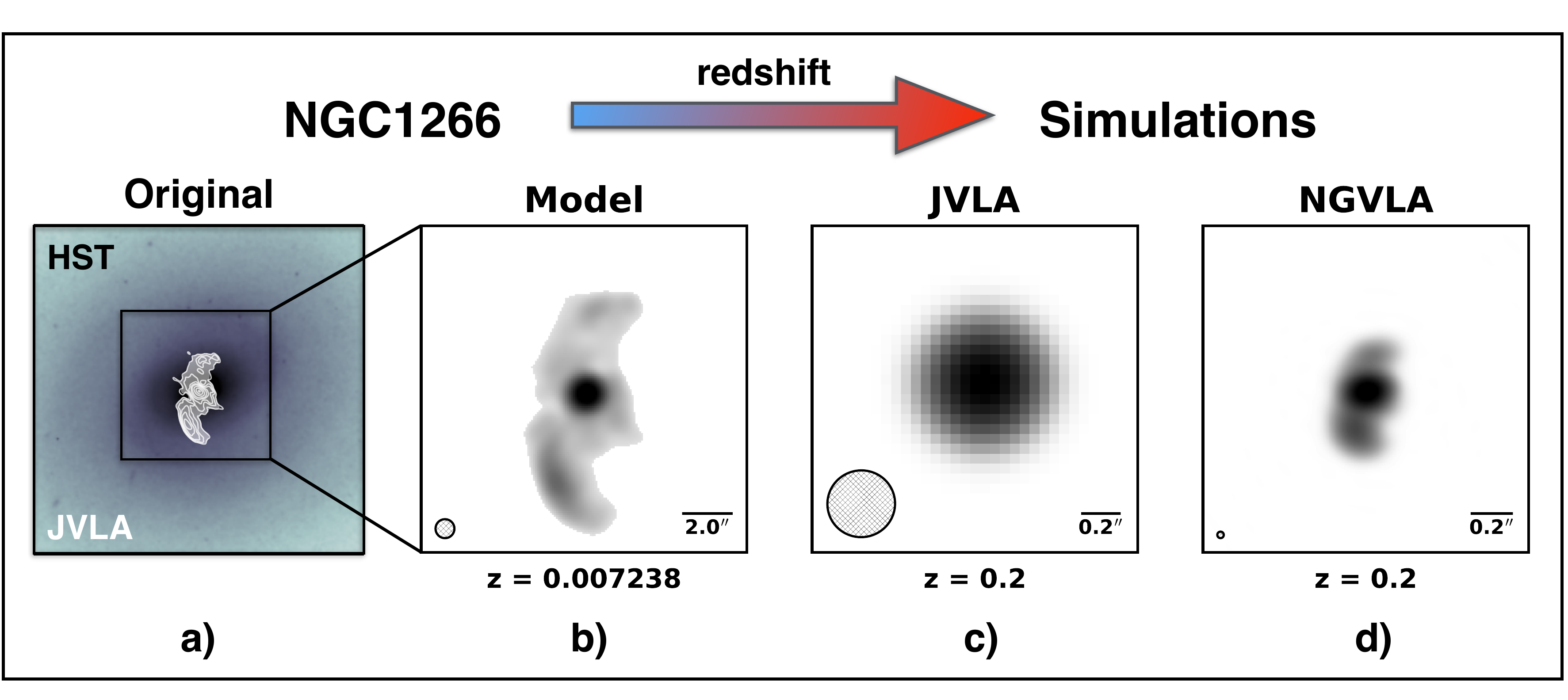}
\caption{Simulated VLA and ngVLA images of a redshifted analog to the nearby jet-driven feedback candidate NGC\,1266.  {\bf a)} A near-infrared Hubble Space Telescope image (WFC3 F140W) from the Hubble Legacy Archive is shown in the background colorscale and the VLA A-configuration radio contours at 5~GHz are overlaid in white.  The angular resolution of the radio data is $\theta_{\mathrm{FWHM}} = 0.75^{\prime \prime}$ and the extent of the radio jets from end to end is 9.5$^{\prime \prime}$ (1.4~kpc).  The NGC\,1266 VLA data were originally published in \citet{nyland+16}.  {\bf b)} Model radio image of NGC\,1266 based on the original image shown in panel a) and shown at the true redshift of the source of $z = 0.00728$. {\bf c)}  Simulated VLA A-configuration map of an NGC\,1266 analog redshifted $z = 0.2$ at 5~GHz with $\theta_{\mathrm{FWHM}} = 0.34^{\prime \prime}$.  {\bf d)} Simulated map of an NGC\,1266 analog at $z = 0.2$ as it would appear if imaged with the ngVLA at 5~GHz.  The angular resolution is $\theta_{\mathrm{FWHM}} = 0.03^{\prime \prime}$.
\\}
\label{fig:NGC1266_sim}
\end{figure*}

\begin{figure*}[t!]
\centering
\includegraphics[clip=true, trim=0.05cm 7cm 0.05cm 4cm, height=0.4\textwidth]{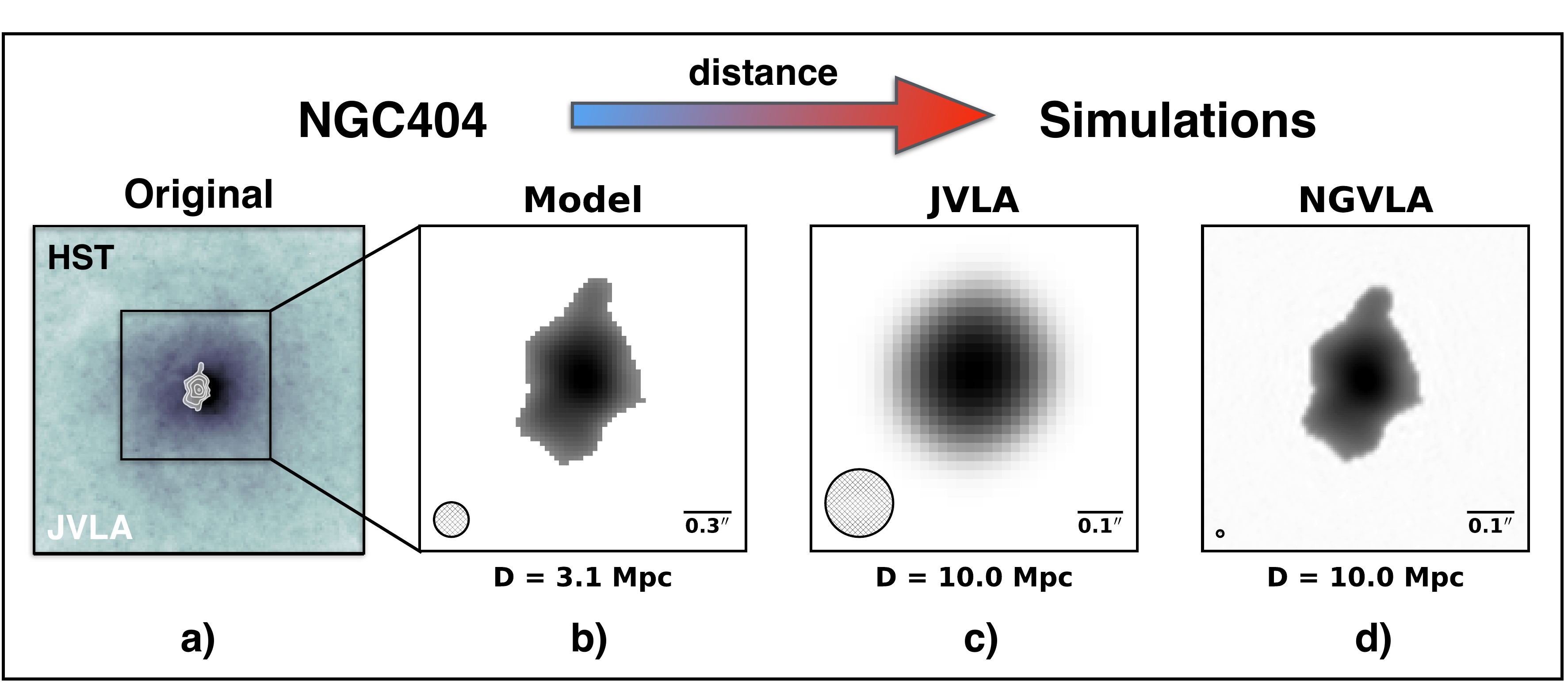}
\caption{Simulated VLA and ngVLA images of an analog to the jetted AGN hosted by the nearby dwarf galaxy NGC\,404 as it would appear if it were $\sim$3X more distant.  {\bf a)} An {\it HST} optical image is shown in the background colorscale and the VLA $Ku$-band (15~GHz) radio contours are overlaid in white.  The angular resolution of the radio data is $\theta_{\mathrm{FWHM}} = 0.20^{\prime \prime}$ and the extent of the radio jets from end to end is 1.13$^{\prime \prime}$ (17~pc).  The NGC\,404 VLA data were originally published in \citet{nyland+17}.  {\bf b)} Model radio image of NGC\,404 based on the original data shown in panel a) and shown at the true source distance of $D = 3.1$~Mpc. {\bf c)}  Simulated VLA A-configuration map of an NGC\,404 analog shifted to a distrance of $D = 10.0$~Mpc at 15~GHz with $\theta_{\mathrm{FWHM}} = 0.15^{\prime \prime}$.  {\bf d)} Simulated map of an NGC\,404 analog at $D = 10.0$~Mpc as it would appear if imaged with the ngVLA at 15~GHz.  The angular resolution is $\theta_{\mathrm{FWHM}} = 0.014^{\prime \prime}$.
\\}
\label{fig:NGC404_sim}
\end{figure*}

\subsubsection{1 kpc-scale Jet}
In Figure~\ref{fig:NGC1266_sim}, we present our VLA and ngVLA simulations of a jetted low-luminosity AGN with properties similar to NGC\,1266, an example of a relatively ``normal,'' nearby ($D = 29.9$~Mpc) galaxy with strong evidence for AGN-driven feedback produced by its low-power ($L_{1.5\,\,\mathrm{GHz}} \sim 10^{22}$~W~Hz$^{-1}$), kpc-scale radio jet.  NGC\,1266 hosts a multi-phase outflow in conjunction with a reduced star formation efficiency in the vicinity of the AGN.  A number of studies over the past several years support a scenario in which the outflow is driven at least in part by the injection of turbulent energy from the radio jet \citep{alatalo+11, alatalo+14, alatalo+15, davis+12, nyland+13}.  
This galaxy thus highlights the importance of turbulence-driven quenching caused by AGN feedback associated with a low-power jet hosted by a galaxy with a stellar mass of $\log(M_{*}/\mathrm{M}_{\odot}) \sim 11.56$, which is comparable to that of the Milky Way. However, given the rarity of such objects, it remains unclear as to whether NGC\,1266 is an anomalous system requiring a special set of conditions to exist, or instead represents a more common phase of galaxy evolution awaiting more widespread identification with improved sensitivity and spatial resolution \citep{nyland+16}.

Figure~\ref{fig:NGC1266_sim} demonstrates the importance of angular resolution for identifying radio jets with kpc-scale extents that may be interacting substantially with the ISM of their hosts.  If NGC\,1266 were moved from its true redshift of $z = 0.007238$ out to $z = 0.2$, its radio jet would be unresolvable by the VLA at 5~GHz -- even when observed in the A configuration ($B_{\mathrm{max}} = 36$~km).  With a factor of ten increase in maximum baseline length, the ngVLA would be able to easily resolve the morphology of a kpc-scale radio jet at $z = 0.2$ at 5~GHz.  Higher-frequency ngVLA continuum observations would yield even higher angular resolution images better suited for detecting the morphological signatures of kpc-scale jets in more distant galaxies.  However, if the jets are characterized by optically-thin synchrotron emission with a radio spectral index of $\approx -0.7$, their intrinsically fainter fluxes at higher frequencies will significantly inflate the on-source integration time needed to detect the emission.

We emphasize that prospects for spatially resolving NGC\,1266 analogs out to even a modest redshift of $z \sim 0.2$ with the current VLA are poor. The highest angular resolution that the VLA can achieve, namely $Q$ band (40$-$50 GHz) observations in A-configuration ($\theta_{\mathrm{FWHM}} \sim 40$~mas), only begins to approach the resolution of the ngVLA at 5~GHz. The steep radio spectral index of NGC\,1266 of $\alpha = -0.79 \pm 0.02$ \citep{nyland+13}, as well as the substantially higher system temperature and correspondingly poorer sensitivity of the high-frequency VLA receivers, would render $Q$-band observations of the redshifted NGC\,1266 analog infeasible. Furthermore, the possibility of an increasing contribution to the continuum flux by thermal emission in the $40-50$~GHz frequency range \citep{condon+92} could lead to additional complications in the scientific interpretation of the data.  Thus, the high angular resolution of the ngVLA (tens of milliarcseconds) in the centimeter-wave regime will serve as a unique but essential tool for resolving the morphologies of kpc-scale radio jets hosted by gas-rich galaxies.  

\subsubsection{10~pc-scale Jet}
\label{sec:NGC404_sims}
We present simulations of a confined jet with an extent of $\sim$10~pc analogous to the source residing in the nucleus of the nearby ($D=3.1$~Mpc), dwarf S0 galaxy NGC\,404 in Figure~\ref{fig:NGC404_sim}.  The central engine of this galaxy is weakly active and powered by a MBH with a mass upper limit from stellar dynamical modeling of $M_{\mathrm{BH}} < 10^5$~M$_{\odot}$ \citep{nguyen+17}, making it the current record holder for the central MBH with the lowest dynamically-constrained mass.  As mentioned in Section~\ref{sec:dwarf_galaxies}, multiwavelength evidence for shocks in the center of NGC\,404 associated with the confined jet raise the intriguing possibility of localized AGN feedback caused by the interaction between the jet and the nuclear molecular gas \citep{nyland+17}.  

The simulations shown in Figure~\ref{fig:NGC404_sim} are based on previous VLA A-configuration observations at 15~GHz.  To produce the input model, we scaled the image pixel size such that it corresponds to an NGC\,404 analog located at 10~Mpc, or $\sim$3X further than the true distance to NGC\,404.  Figure~\ref{fig:NGC404_sim} clearly indicates that the VLA cannot spatially resolve the morphology of a 10~pc jet at a distance of 10~Mpc in the A-configuration.  However, the ngVLA is able to successfully image the extended structure of the $\sim$10-pc-scale jet given the same input model and observing frequency.   
The identification of NGC\,404 analogs in other dwarf or low-mass galaxies may therefore offer new insights into the MBH occupation fraction in a currently poorly-constrained galaxy mass regime and also provide new insights into AGN feedback physics associated with MBHs with masses below 10$^6$~M$_{\odot}$ (e.g., \citealt{volonteri+10, reines+16b}).

\begin{figure*}[t!]
\centering
\includegraphics[clip=true, trim=1.5cm 0.1cm 2.5cm 1cm, height=0.77\textwidth]{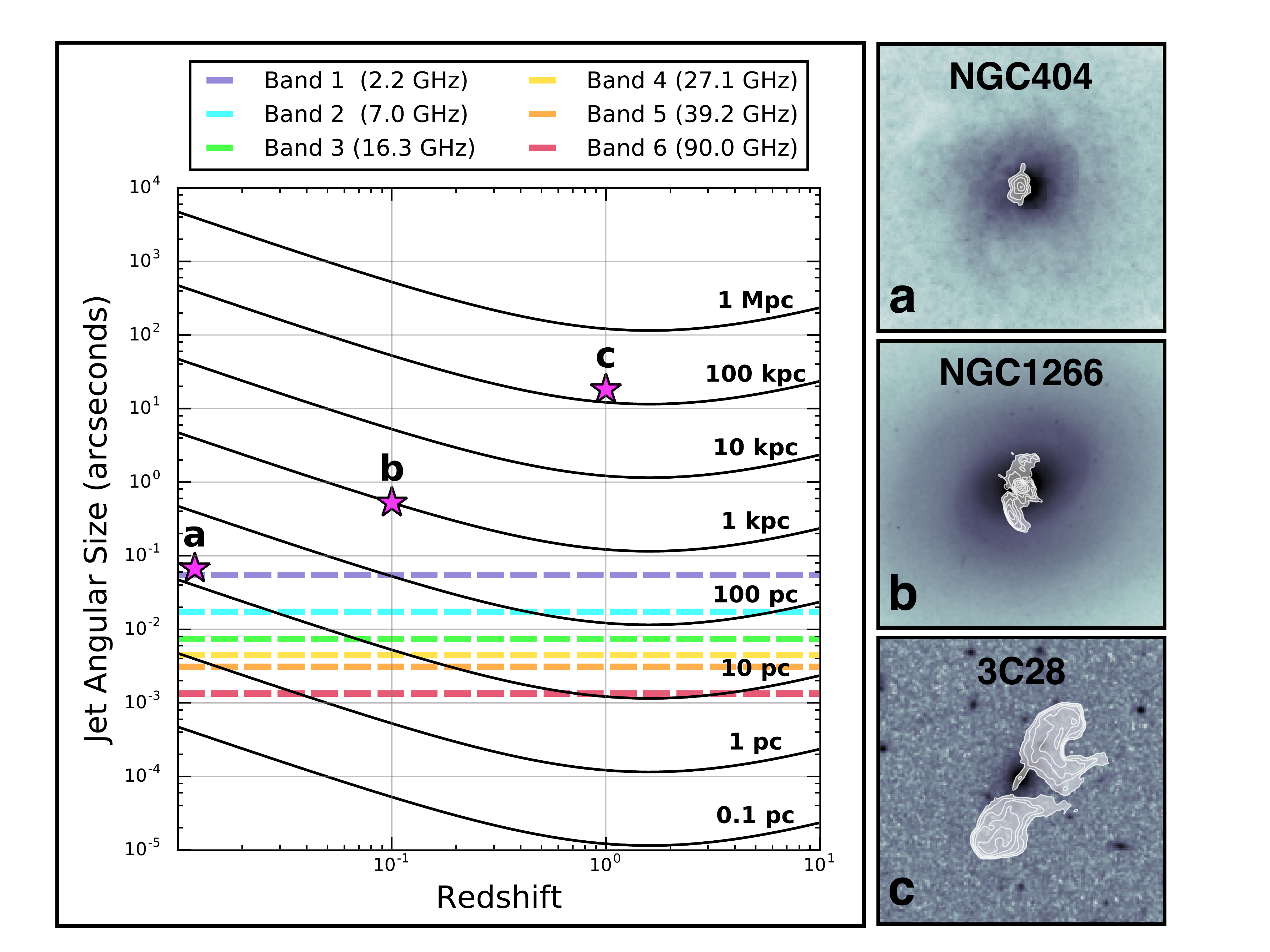}
\caption{Observed jet angular size as a function of redshift.  The black solid lines trace the redshift dependence of the angular extent of a jetted AGN for intrinsic jet sizes (measured from end to end along the major axis of the radio source) ranging from 0.1~pc to 1~Mpc.  The maximum angular resolution of the ngVLA (estimated by $\lambda$/$B_{\mathrm{max}}$, where we assume $B_{\mathrm{max}}$ = 513~km) at the center of each of the ngVLA bands as defined in \citet{Selina+17} is denoted by the dashed colored lines.  The magenta stars highlight the position on the diagram of three  representative cases based on the sources for which simulations were performed in Section~\ref{sec:simulations}: {\bf a)} the dwarf galaxy NGC\,404 with a jet extent of 10~pc, {\bf b)} the jet-driven feedback host NGC\,1266 with a jet extent of 1~kpc, and {\bf c)} the radio galaxy 3C28 with a jet extent of 150~kpc.  The redshifts of the representative sources correspond to those of the simulated ngVLA maps at $z \approx 0$ ($D = 10$~Mpc), $z = 0.2$, and $z = 1.0$, respectively. The thumbnails shown to the right of the main figure illustrate the radio morphologies of the three representative sources at their true redshifts and are described in detail in the far-left panels of Figures~\ref{fig:3C28_sim}, \ref{fig:NGC1266_sim}, and \ref{fig:NGC404_sim}. 
\\}
\label{fig:jet_size_z}
\end{figure*}

\subsection{Observational Implications}
\label{sec:obs_imp}
The simulations presented in Section~\ref{sec:simulations} were designed to explore the ability of the $uv$-coverage of the antenna configuration shown in Figure~\ref{fig:config} to adequately capture the range of spatial scales present in our input source models.  Sensitivity considerations were not included in these simulations.  Here, we discuss the observational implications for future ngVLA studies of radio jets given the surface brightness sensitivity of the ngVLA. 

In Figure~\ref{fig:jet_size_z}, we illustrate the redshift dependence of the observed angular jet extent for a wide range of radio jet size scales ranging from sub-parsec jets to giant radio galaxies with Mpc-scale lobes.  The maximum angular resolution (defined as $\theta_{\mathrm{max}} = \lambda/B_{\mathrm{max}}$) for each of the proposed ngVLA observing bands (assuming $B_{\mathrm{max}} \approx 500$~km) is also highlighted.  
Figure~\ref{fig:jet_size_z} clearly indicates that the ngVLA will not be able to resolve the morphologies of jets with extents in the range of $0.1$~pc $\lesssim d_{\mathrm{jet}} \lesssim 1$~pc at any redshift.\footnote{While the two highest frequency bands of the ngVLA with proposed central frequencies of 39.2 and 90.0~GHz could in principle spatially resolve a 1~pc-wide jet at low redshift, such observations would be impractical for sources with steep, optically-thin synchrotron spectral indices that would be intrinsically very faint in this frequency range.}  Resolved imaging of jets with extents of tens of parsecs would be feasible over the redshift range of $z = $ 0.1 to 1, and jets spanning 100~pc or larger could in principle be spatially resolved in great detail by the ngVLA.  

However, with its fixed maximum baselines out to a few hundred km, the surface brightness sensitivity (the point source sensitivity per beam solid angle) of the ngVLA will make it difficult or impossible to image sources with large angular extents relative to the size of the synthesized beam.  This is because the angular resolution scales linearly with maximum baseline length, but surface brightness sensitivity scales as the square of the beam solid angle (given constant point source sensitivity).  Thus, given the antenna configuration considered in our study, the ngVLA is poorly optimized for imaging extended, diffuse sources.  

We summarize the feasibility of our ngVLA and VLA simulated radio AGN observations in Table~\ref{tab:obs_summary}.  For the 3C28 analog simulation described in Section~\ref{sec:sims_100kpc_jet}, the $\sim$ 100~kpc-wide radio jets have an angular extent of about $18^{\prime \prime}$ at $z = 1$.  At the angular resolution of our simulated ngVLA observation at 6~GHz of $\theta_{\mathrm{FWHM}} = 28$~mas, the total flux of the source will be spread over $> 4.0 \times 10^5$ synthesized beams.  Given the redshifted total flux of 3C28 (see Table~\ref{tab:sim_summary}), a $5\sigma$ detection with a resolution of $28$~mas would require an on-source integration time of $5.83 \times 10^4$~ hours ($\sim$6.65 years).  Thus, the surface brightness sensitivity of the ngVLA on the longest baselines is insufficient for imaging radio jets spanning hundreds of kpc at the angular resolution shown in the far-right panel of Figure~\ref{fig:3C28_sim}. 

\begin{deluxetable*}{ccccCCCCCCC}[t!]
\tablecaption{Feasibility of Simulated Radio AGN Observations  \label{tab:obs_summary}}
\tablecolumns{11}
\tablewidth{0pt}
\tablehead{
\colhead{Source} & \colhead{Band} & \colhead{$\theta_{\mathrm{max}}$}  & \colhead{$\theta_{\mathrm{sim}}$}  & \colhead{$N_{\mathrm{sim}}$} & \colhead{$FOM_{\mathrm{sim}}$} &  \colhead{$t_{\mathrm{sim}}$} &  \colhead{$\theta_{\mathrm{tap}}$} & \colhead{$N_{\mathrm{tap}}$} & \colhead{$FOM_{\mathrm{tap}}$} & \colhead{$t_{\mathrm{tap}}$}\\
\colhead{} & \colhead{} & \colhead{(mas)} & \colhead{(mas)} & \colhead{} & \colhead{($\mu$Jy~beam$^{-1}$)} & \colhead{(hr)} & \colhead{(mas)} & \colhead{}  & \colhead{($\mu$Jy~beam$^{-1}$)} & \colhead{(hr)}\\
\colhead{(1)} & \colhead{(2)} & \colhead{(3)} & \colhead{(4)} & \colhead{(5)}  & \colhead{(6)}  & \colhead{(7)}  & \colhead{(8)}  & \colhead{(9)} & \colhead{(10)} & \colhead{(11)}\\
}
\startdata
\hline
ngVLA Simulations \\
\hline
3C28 & 2 & 20 & 28 & $4.24 \times 10^{5}$ & $5.91 \times 10^{-4}$ & $5.83 \times 10^{4}$ & 340 & $2.82 \times 10^{3}$ &  $8.72 \times 10^{-2}$  & 10.03\\
NGC\,1266 & 2& 24 & 30 & $2.16 \times 10^{2}$  &  $3.12 \times 10^{-2}$ & 20.87 & 38 & $1.34 \times 10^{2}$ & $5.01 \times 10^{-2}$ & 8.11\\
NGC\,404 & 3 & 8 & 14 & $6.26 \times 10^{2}$ & $8.92 \times 10^{-3}$ & $2.29 \times 10^{2}$ & 28 & $1.56 \times 10^{2}$ & $3.56 \times 10^{-2}$ & 14.37 \\
\hline
VLA Simulations \\
\hline
3C28 & $C$ & 283 & 340 & $2.82 \times 10^{3}$  & $8.72 \times 10^{-2}$ & $5.39 \times 10^{4}$ & \nodata & \nodata & \nodata & \nodata  \\
NGC\,1266 & $C$ & 340 & 340 &  \nodata & 4.01 & 1.00 & \nodata & \nodata & \nodata & \nodata   \\
NGC\,404 & $Ku$ & 113 & 150 & \nodata & 1.02 & 12.75 & \nodata & \nodata & \nodata & \nodata  \\
\enddata
\tablecomments{Column 1: Source name.  Column 2: Band name of simulated observations.  Bands 2 and 3 refer to the ngVLA bands centered at 7.0 and 16.3~GHz with bandwidths of 8.7 and 8.4~GHz, respectively, as defined in \citet{Selina+17}.  Bands $C$ and $Ku$ refer to the VLA bands covering the frequency ranges of $4-8$ and $12-18$~GHz, respectively.  Column 3: Maximum angular resolution, defined as $\theta_{\mathrm{max}} = \lambda/B_{\mathrm{max}}$.  For the ngVLA simulations, we assume $B_{\mathrm{max}} = 513$~km and for the VLA simulations we assume the maximum baseline length in the A configuration of 36.4~km.  Column 4: FWHM of the major axis of the synthesized beam from our simulated image.  Column 5: Approximate number of synthesized beams subtended by the source defined as $N_{\mathrm{sim}} = (\theta_{\mathrm{jet}}^{\prime})^2/(\theta_{\mathrm{max}})^2$, for the $\theta_{\mathrm{jet}}^{\prime}$ value of the source defined in Table~\ref{tab:sim_summary}.  Column 6: Figure of merit of the simulated observations defined as $FOM_{\mathrm{sim}} = F_{\mathrm{Total}}^{\prime}/(5 \times N_{\mathrm{sim}})$, where $F_{\mathrm{Total}}^{\prime}$ is the redshifted source flux given in Column 11 of Table~\ref{tab:sim_summary}.  Column 7:  On-source integration time (not including any calibration overheads) needed to reach $FOM_{\mathrm{sim}}$ from Column 6 using the performance estimates given in \citet{Selina+17} for the ngVLA and the exposure calculator for the VLA.  Column 8:  Synthesized beam FWHM of the Gaussian $uv$-taper applied to the simulated data.  Column 9: Same as Column 5, but for the tapered synthesized beam size defined in Column 8.  Column 10: Same as Column 6, but for the number of synthesized beams subtended by the source given in Column 9.  Column 11: Same as Column 7, but for the $FOM$ given in Column 10.  
}
\end{deluxetable*}

The final sensitivity and angular resolution of an interferometric image ultimately depend on a number of parameters, including the $uv$-coverage (antenna configuration), the density weighting of the visibilities (e.g., natural, uniform, or Briggs robust weighting), and the level of Gaussian tapering applied during imaging to suppress high spatial frequencies (e.g., \citealt{cotton+17}).  It is thus possible to increase the surface brightness sensitivity of the simulated 3C28 image by applying a Gaussian taper to the data during imaging, which increases the synthesized beam size at the cost of point source sensitivity.  Using the weighted sensitivity estimates from \citet{Selina+17}, we estimate the on-source integration time needed to detect the redshifted 3C28 analog with the ngVLA if the data were tapered down to the maximum resolution of the VLA A-configuration at 6~GHz ($\sim$340~mas).  We find that such an observation would be feasible, requiring only $\sim$10 hours of observing time with the ngVLA.  We note that observations with the same angular resolution using the VLA would not be feasible given its smaller collecting area and the limited bandwidth range of the $C$-band receiver compared to the ngVLA.  

Table~\ref{tab:obs_summary} also summarizes the feasibility of the redshifted NGC\,1266 and NGC\,404 simulated VLA and ngVLA observations of jets with angular sizes of $\sim$1~kpc to $\sim$10~pc, respectively.  The VLA would be able to detect the emission from both of these simulated sources in a reasonable amount of on-source integration time (defined here arbitrarily as $t_{\mathrm{sim}} < 15$~hrs), however, the VLA lacks sufficiently long baselines to spatially resolve the morphologies of these sources.  This is problematic since the low luminosities of these sources (particularly in the case of NGC\,404, which has a 15~GHz luminosity of only $\sim$3.4$\times 10^{17}$~W~Hz$^{-1}$) make it difficult to distinguish their origins from nuclear star formation without additional morphological constraints to help identify them as being powered by MBH accretion (e.g., \citealt{nyland+13, nyland+16, nyland+17}).  The ngVLA, on the other hand, would be capable of detecting and spatially resolving these sources within a reasonable amount of time given a modest amount of tapering ($20-50$\% lower resolution than the ngVLA simulations shown in Figures~\ref{fig:NGC1266_sim} and \ref{fig:NGC404_sim}, respectively).

Thus, studies of radio jets with intrinsic extents of a few pc to a few kpc will fully utilize the unique combination of angular resolution, collecting area, and frequency coverage of the ngVLA over a wide range of redshifts.  This range of angular scales is well-matched to the expected extents of young radio AGNs and, combined with the frequency range of the ngVLA, further emphasizes the suitability of the ngVLA for studies of jet-ISM feedback associated with young and/or lower-power radio AGNs.    Observations of more extended radio jets that may be engaged in feedback on intergalactic or intracluster medium scales will be possible with suitable combinations of weighting and $uv$-tapering.  However, the poor surface brightness sensitivity on the longest ngVLA baselines will prevent radio AGN studies of extended sources at the maximum resolution of the array.  


\section{Computational Challenges}
\label{sec:algorithms}
The roughly order-of-magnitude increase in the sensitivity and maximum baseline lengths of the ngVLA compared to the VLA, as well as prospects for ultra-wide bandwidth ratio receivers to accommodate the unprecedented frequency range proposed for the ngVLA, will revolutionize our understanding of the radio sky, but not without substantial computational challenges. Here, we briefly discuss some of the most important computational considerations for the ngVLA that will have a significant impact on continuum studies of AGNs. For a more in-depth assessment of the algorithmic and computational challenges to be faced by the SKA and its pathfinders -- which like the ngVLA consist of large numbers of antennas distributed over wide geographical areas -- we refer readers to \citet{norris+13}. 

\subsection{Image Dynamic Range}
An important challenge to overcome, particularly in the lower-frequency regime of the ngVLA, will be producing images of sufficiently high dynamic range to detect radio sources that are extremely faint or characterized by complex morphologies.  The near order-of-magnitude gain in continuum sensitivity compared to the VLA will mean that images of sources with fluxes on the order of a few mJys will begin suffering from dynamic range effects.  Additional factors that will ultimately influence the dynamic range of ngVLA images -- and the algorithmic requirements for forming them -- include observing bandwidth, field-of-view, instrumental errors, antenna pointing robustness, and interference from the ionosphere.

\subsection{High Bandwidth Ratios}
Wide-band imaging will be necessary for characterizing radio AGN spectral indices and performing spectral aging analyses.  However, generating images over a wide-range of frequencies with the 2:1 bandwidth ratio receivers of the VLA at $L$ (1--2~GHz), $S$ (2--4~GHz), and $C$ (4--8~GHz) band is already known to lead to dynamic-range effects that become significant around dynamic ranges of $\sim$1000.  In this regime, techniques such as multifrequency synthesis, or multi-term multi-frequency synthesis (MTMFS; \citealt{rau+11}) for extended sources, must be implemented.  Receivers with even wider bandwidth ratios of $\sim$4:1 have been proposed for the ngVLA \citep{Selina+17}.  Such ultra-wide-bandwidth receivers would therefore be expected to be even more susceptible to first-order dynamic range effects, and may require the development of new wide-band imaging algorithms.

subsection{Spectral and Time Resolution}
The proposed ultra-wide-bandwidth receivers for the ngVLA \citep{Selina+17}, coupled with the large field of view of the 18m antennas, requires high spectral channel resolution to avoid image degradation due to bandwidth smearing.  For wide-field continuum imaging, the maximum channel width requirement decreases with observing frequency, leading to narrower channel widths and thus more computationally expensive requirements in the lowest frequency ngVLA bands.  For the case of a Gaussian bandpass with circular Gaussian tapering (Equation 18-29 in \citealt{bridle+99}), we find that a channel width of $\Delta \nu_{\mathrm{chan}} \approx 30$~kHz for ngVLA Band~1 (1.2 to 3.9~GHz) would lead to a reduction in peak intensity for sources at the half-power beamwidth (HPBW) of the primary beam\footnote{For the primary beam, the half-power beamwidth for observations at wavelength $\lambda$ is $\theta_{\mathrm{HPBW}} = 1.2 \lambda/D$, where $D$ is the antenna diameter.} (PB) to 10\%.  More stringent requirements on the acceptable level of source amplitude loss at the PB HPBW would require observations with even narrower channel widths.  

The longest baselines of the ngVLA of up to several hundred km also place constraints on the minimum correlator integration time during widefield imaging to avoid substantial time-average smearing, which degrades the image quality and leads to a reduction in observed source amplitude.  To limit amplitude losses to 10\% at the PB HPBW for the maximum baseline length of 513~km considered in this study, we estimate a maximum correlator integration time of $\sim$0.5 seconds (assuming circular $uv$-coverage with Gaussian tapering as in Equation 18-43 in \citealt{bridle+99}).  
Combined with the inherently large number of baselines for a 300 element array, the high spectral and temporal resolution that will be necessary for producing high-fidelity images over the full field-of-view of the ngVLA will create new challenges for data storage and processing that must be addressed in the decade leading up to the construction of the ngVLA.  

\subsection{Instrumental Effects}
At high dynamic ranges (e.g., > 10$^4$), or for cases where off-axis sources in wide-field images are of interest (either scientifically in the case of wide-area surveys, or for the removal of sidelobes affecting the pointing center that originate from distant, bright sources), instrumental effects become important.  These effects include time variations in the PB as well as instrumental polarization leakage.  Imaging fields including sources with fluxes in the range of tens of mJys will result in dynamic ranges near 10$^4$, and the ability to produce high-fidelity images of such sources with the ngVLA will require the use of cutting-edge imaging techniques such as wide-band ``full-mueller matrix'' imaging (e.g., \citealt{jagannathan+17}), which account for the frequency and time dependence of the antenna PB as well as polarization leakage.

Accurate antenna pointing solutions will be important for high-dynamic-range imaging over the full frequency range of the ngVLA, particularly for mosaics and wide-field images.  Antenna pointing accuracy of the ngVLA will be complicated by the large number of antennas in the array, and therefore, techniques such as pointing self calibration (e.g., \citealt{bhatnagar+04}), may be necessary.  This technique, when coupled with the imaging strategies described above, would account for direction-dependent instrumental gains during the imaging major cycle. 


\subsection{Data Volume and Processing}
Imaging AGNs in the lower-frequency bands of the ngVLA will require large images of the full field-of-view (and possibly beyond it if bright sources are present in the PB sidelobes) of the 18m ngVLA antennas.  The resulting continuum images may span tens of GB and originate from datasets of tens or hundreds of TB.  Cubes from spectral line observations may be even more unwieldly, spanning hundreds of GBs or even reaching 1~TB or more.  Observations of cold gas indicating fast outflows driven by powerful AGNs, which may span hundreds or even thousands of km~s$^{-1}$ (or more narrow lines that require ultra-high spectral resolution for the identification of shifted features potentially associated with inflow/outflow), will present particularly challenging cases.  Thus, the ability to pursue exciting and new studies of radio-jet feedback using the ngVLA demands imaging algorithms that can handle memory and computations in an efficient manner.  The development of ``scalable algorithms'' that employ parallel or distributed schemes will therefore be an integral part of both preparing for the ngVLA and supporting its operations once functional.

\section{Multiwavelength Synergy}
\label{sec:synergy}
There is no other radio telescope in the world -- existing or planned -- that will produce images of comparable sensitivity and angular resolution over the wide observing frequency range of the ngVLA.  As discussed in Section~\ref{sec:science}, the unique combination of features of the ngVLA will facilitate a number of advancements in the field of AGN science.  However, improving our understanding of jet-ISM feedback, the elusive population of AGNs residing in low-mass and/or gas-rich galaxies, and the connection between AGNs and galaxy evolution cannot be accomplished with ngVLA observations alone.  Thus, in this section we consider synergies between the ngVLA and existing/future telescopes operating in a wide range of observational regimes.

\subsection{ALMA and the VLA}
Currently operating radio facilities, such as the VLA and ALMA, are beginning to perform pioneering studies of radio jets and the circumnuclear gas in the vicinity of AGNs  \citep{garcia-burillo+14, morganti+15b, lonsdale+16, russell+17}.  Indeed, the full scientific potential of the VLA and ALMA for improving our understanding of jet-ISM feedback has yet to be reached.  Future VLA and ALMA observing programs studying the continuum and cold gas conditions in large samples of gas-rich galaxies harboring jetted low-luminosity AGNs will undoubtedly make important contributions to our understanding of how jets affect the ambient gas conditions.  However, these studies will face a variety of challenges due to key differences between the VLA and ALMA, such as frequency coverage, field-of-view, survey speed, angular resolution\footnote{Depending on the respective observing bands of the VLA and ALMA observations in question, their angular resolutions may be comparable when the VLA is in an extended configuration and ALMA is in a more compact one.}, and geographical location.

The ngVLA will offer the ability to probe continuum emission over an ultra-wide frequency range in the cm and mm regimes, while also providing an improvement in sensitivity compared to both ALMA and the VLA. Where continuum observations with ALMA will mostly trace thermal dust emission, the ngVLA will be sensitive to synchrotron emission from AGN and star formation. Additionally, the ngVLA will be better suited for deep continuum and CO studies than ALMA over the range of overlapping frequencies between these two telescopes.   

ALMA can only trace the ground-transition of CO ($\nu_{\rm rest}\,=\,115.2712$ GHz) at low redshifts, although its future Band 1 (35--50 GHz) receiver may extend this to $z$\,$\sim$\,2.3. The ngVLA will be sensitive to CO(1--0) out to much higher redshifts. CO(1--0) has a lower critical density and is less dependent on the excitation conditions of the gas than higher-$J$ lines. Therefore, CO(1--0) is the most robust tracer of the overall molecular gas content of galaxies, including any widespread or sub-thermally excited component. The ngVLA is the {\it only} instrument that has the frequency range and sensitivity to trace the ground-transition of CO across almost the entire history of the universe, complementing observations of the high-$J$ transitions with ALMA. Deep observations of low-$J$ CO transitions with the ngVLA will be able to detect the kinematic signatures of outflows traced by faint molecular gas emission both in new classes of galaxies in the low-redshift universe (e.g., nearby dwarf galaxies harboring low-mass AGNs) and in high-redshift galaxies in the process of assembly.  

In addition to the substantial improvement in sensitivity, the accessibility of the low-$J$ CO lines within the frequency range of the ngVLA will help decrease observational demand in this regime for ALMA.  This would free-up additional time for ALMA to perform observations of denser gas tracers uniquely accessible in ALMA bands 4 through 10 ($125-950$~GHz).  Ultimately, the synergistic combination of cm-wave continuum observations and lower-density molecular gas data from the ngVLA with higher-density gas measurements from ALMA will provide vital insights into the chemistry and physics of jet-ISM feedback in unprecedented detail.  

\subsection{SKA}
\label{sec:synergy_SKA}
The advent of the SKA, a new radio telescope currently under development that will ultimately have a collecting area of one square kilometer,  is expected to complement the ngVLA by producing transformative new science over a wide range of topics \citep{carilli+04, carilli+15, braun+15}.  The SKA project development has been divided into two phases, with phase 1 construction expected to begin over the next few years.  Two instruments will be built during phase 1: SKA1-LOW and SKA1-MID \citep{dewdney+15}.  SKA1-LOW will operate from 50 to 350~MHz and have a collecting area of 0.4~km$^2$ with maximum baselines of $\sim$65~km.  The SKA1-MID will consist of 197 antennas (133 $\times$ 15m SKA1 + 64 $\times$ 13m MeerKAT dishes) with maximum baselines of 150~km and operate from 0.35--14~GHz, thus overlapping with the proposed frequency range of the ngVLA (1 to 116~GHz).  In this study, unless otherwise noted, when referring to the observational capabilities of the SKA we are referring to the SKA1 phase only.  

The key science drivers of the SKA encompass many important areas of research related to AGN physics and the connection between MBHs and galaxies in the context of galaxy formation and evolution.  As detailed in numerous recent white papers, the SKA will directly probe jet-ISM feedback through observations of neutral hydrogen absorption against radio jets, constrain radio AGN duty cycles through spectral aging studies, and provide a census of the radio AGN population at low luminosities and high redshifts \citep{agudo+15, kapinska+15, kharb+16, mcalpine+15, orienti+15, prandoni+15, smolcic+15}.  Thus, there is indeed significant overlap between the scientific questions that the SKA and the ngVLA will address regarding AGNs (see Section~\ref{sec:science}), but we emphasize that this overlap is synergistic in nature.  

\subsubsection{Cold Gas} The ngVLA and SKA will both probe the 21~cm line and trace jet-ISM feedback in the atomic phase through \hi\ absorption at low redshifts. While only the SKA can observe the neutral H{\tt I} content of the universe out to $z \sim 1$ to $2$, which approaches the peak epoch of cosmic galaxy assembly, it cannot target the higher frequency tracers of molecular gas. The ngVLA will be uniquely suited for studying the molecular gas content of galaxies over a wide range of redshifts by targeting the lower$-J$ transitions of CO.

\subsubsection{Broadband radio SEDs} The ngVLA and SKA will also both allow broadband studies of radio SEDs over their respective wide frequency ranges, thus providing the temporal information from spectral aging models necessary for interpreting the evolutionary stage of the source. The SKA will be able to detect emission associated with the fading radio lobes, tracing the regime of radio AGNs with older spectral ages (``AGN archaeology''; Morganti 2017). In Figure~\ref{fig:JP}, we show that the 1 to 116 GHz observing range of the ngVLA is advantageous for studies of low-redshift radio AGNs, or intermediate redshift sources that are very young. Such young sources are often embedded in dense environments, and are known to drive multi-phase gas outflows \citep[e.g.,][]{holt+08,gereb+15}. Broadband radio continuum surveys with both the ngVLA and SKA will therefore be needed to construct a complete picture of the life cycles of radio AGNs and their connection to galaxy evolution.  

\subsubsection{High-redshift radio AGNs} Section~\ref{subsec:AGNatHighz} highlighted the importance of studying radio AGNs at higher redshifts ($z\gtrsim4$) where there is a dearth of observational data. Steep radio spectral index selection may be an important technique for identifying high-redshift radio galaxies  \citep{rottgering+94,tielens+79,blumenthal+miley79}. Such radio sources will be brightest at the lowest frequencies of the SKA, enabling the possibility of efficient identification of high-redshift candidates in survey data. By identifying these candidates with SKA1-LOW surveys and following them up with higher-resolution observations with the ngVLA, we can significantly increase our knowledge of high-redshift galaxies, and advance our understanding of their role in galaxy evolution. 

\subsubsection{Polarization} In addition to the science topics described above, both the ngVLA and SKA will probe the broadband polarimetric properties of AGNs.  As described in \citet{gaensler+15}, full-polarization SKA observations will detect and map the RMs of the faint-source population at high angular resolution, thus providing new insights into the relationships between MBHs and their environments. \citet{gaensler+15} estimate that the SKA1-MID will survey sources with RMs $\sim$ 5~rad~m$^{-2}$ (consistent with the typical RMs of the mJy polarized source population at centimeter wavelengths; \citealt{hales+14, rudnick+14}) out to $z \sim 1$.  However, as discussed in Section~\ref{sec:pol}, sources with extreme RMs of tens of thousands of rad~m$^{-2}$ may be depolarized over the full frequency range of the SKA1-MID.  This class of extreme-RM sources, the physical nature of which is currently  unclear, will thus only be accessible to polarimetric studies with the ngVLA at millimeter wavelengths.  In addition to its sensitivity to extreme Faraday rotation environments, the ngVLA will also have an advantage over the SKA1-MID in terms of reduced beam depolarization as a result of its higher angular resolution.

\subsubsection{Observing Logistics} We also highlight some logistical differences between the ngVLA and SKA.  While the SKA is intended to be operated primarily as a survey instrument, the current vision for the ngVLA is to emphasize P.I.-driven science led by members of the community.  Thus, the ngVLA would be ideally suited for follow-up observations of sources of interest identified in surveys by the SKA or its pathfinders (e.g., \citealt{shimwell+17}) at higher frequency and angular resolution. We note that e-Merlin is an existing P.I.-driven instrument that has similar longest baselines (currently 220~km).  Thus, e-MERLIN is in this sense an ngVLA pathfinder (albeit with much sparser $uv$- and frequency-coverage and lower collecting area).

The location of the ngVLA in New Mexico will allow high-sensitivity studies of key individual sources in the northern hemisphere inaccessible or with poor visibility to the SKA due to its location in the Southern hemisphere.  Examples of northern sources of interest include bright radio galaxies (e.g., Cygnus~A; \citealt{israel+98}) and nearby, low-mass galaxies with candidate jetted low-luminosity AGNs (e.g., NGC\,404; \citealt{nyland+17}) or intermediate-mass black holes (e.g., M31 and M82; \citealt{mezcua+18}, and references therein).

\subsection{Infrared and Optical Telescopes}

\subsubsection{{\it James Webb} Space Telescope}
The upcoming launch of the {\it James Webb} Space Telescope ({\it JWST}) will enrich the astronomical community with a wealth of new data for AGN studies.  Deep, high-resolution imaging with {\it JWST} will spatially resolve the morphologies of the host galaxies of AGNs out to high redshifts, and will prove invaluable to our understanding of the evolutionary link between MBHs and the galaxies in which they reside.  {\it JWST} will accomplish this in a number of ways, including its ability to provide essential diagnostic information to help distinguish between different underlying emission mechanisms, such as star formation and an AGN.  For example, mid-infrared (MIR) imaging will allow star formation/AGN decomposition in lower-luminosity and/or more distant hosts through color diagnostics and SED fitting (e.g., \citealt{lacy+07, donley+12}).  This diagnostic information may then be combined with ngVLA measurements of the radio continuum properties of the source to place constraints on the potential energetic impact of feedback in the poorly understood regime of less powerful and/or high-redshift AGNs.

{\it JWST} will also play an exciting role in direct studies of AGN feedback through the detection of outflows.  In the NIR and MIR, {\it JWST} will observe the ro-vibrational lines of warm/hot molecular H$_{2}$ gas, using integral-field spectroscopy. By comparing the distribution and kinematics of warm/hot H$_{2}$ with that of the cold molecular gas, as traced by the low-$J$ CO lines with ngVLA, the physics of AGN-driven outflows of molecular gas can be studied in detail \citep[see, e.g.,][]{rupke+13,emonts+14,pereira+16}. MIR spectral-line diagnostics of features such as [NeV], the coronal line [SiVI], and polycyclic aromatic hydrocarbons (PAHs) will help distinguish between nuclear engines powered by AGN/star formation out to $z \sim 1.5$ \citep{tommasin+10}.  

\subsubsection{Giant Segmented Mirror Telescopes}
The next generation of optical/NIR Giant Segmented Mirror Telescopes (GSMTs), such as the Thirty Meter Telescope (TMT), the Giant Magellan Telescope (GMT), and the European Extremely Large Telescope (E-ELT), will offer angular resolution on milliarcsecond-scales comparable to that of the upper-end of the frequency range of the ngVLA.  Integral field unit spectroscopy with GSMTs will also provide complementary measurements of warm ionized and molecular gas inflow/outflow associated with MBH fueling and AGN-driven outflows of different gas phases compared to the ngVLA on the same spatial scales (e.g., \citealt{schonell+14,harrison+15,may+16}), and also provide diagnostic information on the dominant underlying excitation mechanism such as an AGN, star formation, and shocks (e.g., \citealt{davis+12}).  Comparison between ngVLA radio continuum observations and optical tracers such as H$\alpha$ can also be used to help determine the relative fractions of thermal and non-thermal radio emission, which will provide new insights on cosmic ray propagation in different galactic environments \citep{tabatabaei+07}. 

\subsubsection{Survey Telescopes}
Dedicated optical and infrared surveys producing large databases of publicly accessible data, such as the Sloan Digital Sky Survey (SDSS; \citealt{york+00}) and the Two-Micron All Sky Survey (2MASS; \citealt{skrutskie+06}), have led to dramatic improvements in our understanding of galaxy evolution by establishing basic galaxy properties (e.g., distances, stellar masses, star formation rates, and nuclear activity levels) suitable for statistical studies.  A new generation of ground- and space-based telescopes will continue this legacy in the decade leading up to the ngVLA.  

At optical wavelengths, the wide field-of-view, high-cadence, and depth of the Large Synoptic Survey Telescope (LSST; \citealt{ivezic+08}) will catalog billions of galaxies, benchmarking their properties and identifying targets for subsequent follow-up at radio wavelengths with the ngVLA.  Furthermore, the high-cadence of the LSST will provide a new avenue for AGN detection and characterization through optical variability measurements (e.g., \citealt{choi+14}).  This novel AGN selection technique will be sensitive to AGNs with low-luminosities, such as those associated with low-mass MBHs, a poorly understood source population that the ngVLA is also well-equipped to explore (Section~\ref{sec:dwarf_galaxies}).  

In the near infrared, space-based observatories, such as the Wide-field Infrared Survey Telescope ({\it WFIRST}; \citealt{spergel+15}), will provide high-angular-resolution imaging comparable to that of the {\it Hubble} Space Telescope, but over a field-of-view that is two orders of magnitude larger.  {\it WFIRST} will identify mildly obscured AGNs that evade optical diagnostic methods, and also probe the rest-frame ultraviolet light of galaxies well into the epoch of reionization.  Together, the combination of the extensive LSST and {\it WFIRST} databases with targeted ngVLA observations will probe the radio properties of galaxies over a broad range of redshifts, thus offering important tests of MBH-galaxy coevolution predictions from cosmological simulations.

\subsection{X-ray Observatories}
The sensitivity and angular resolution of the ngVLA will lead to strong synergies for AGN science with future X-ray observatories, including the Advanced Telescope for High ENergy Astrophysics ({\it Athena}; \citealt{barcons+17}) and {\it Lynx} (formerly {\it X-ray Surveyor}; \citealt{gaskin+16}).  

\subsubsection{{\it Athena}}
{\it Athena} is a fully funded ESA-led mission planned for launch in 2028.  The key features of this telescope are its wide field-of-view and correspondingly fast survey speed, as well as its sensitivity to weak X-ray spectral lines.  
This telescope will offer a significant improvement in collecting area compared to {\it Chandra} and {\it XMM Newton} and moderately high angular resolution with an on-axis point spread function half-energy width of $\sim 5^{\prime \prime}$.  {\it Athena's} two instruments, the wide-field imager and the X-ray integral field unit, feature an impressive field of view of $40^{\prime} \times 40^{\prime}$ and a spectral resolution of up to 2.5~eV, respectively.  

{\it Athena} will address key questions related to the ``hot energetic universe'' theme \citep{nandra+13}, and will thus play an important role in improving our understanding of the formation and growth of MBHs at high redshift and AGN-driven feedback.  In combination with radio continuum observations, X-ray spectroscopic measurements (e.g., abundances, temperatures, shock speeds, etc.) with {\it Athena} will probe the feedback effects of jet-driven shocks associated with kpc-scale radio jets \citep{mingo+15}.  However, we note that {\it Athena's} angular resolution of $\sim 5^{\prime \prime}$ is not an ideal match for that of the ngVLA at most energies.  Nevertheless, its spectroscopic power will be sufficient to model multiple model components in the regions that are spatially confused, providing broad estimates for jet-ISM interactions even when multiple components are spatially unresolved. 

\subsubsection{{\it Lynx}}
The unprecedented sub-arcsecond angular resolution of {\it Chandra} ($<0.5\arcsec$), in combination with continuum VLA observations, has proven to be vital for spatially-resolved studies of the high-energy signatures of jet-driven feedback (e.g., \citealt{wang+11, maksym+17}).  Thus, the combination of ngVLA observations with a high-spatial-resolution X-ray telescope with significantly higher collecting area compared to {\it Chandra} would usher in a new era in our understanding of how AGNs transform the conditions in their ambient environments.  A proposed next-generation X-ray telescope that would meet these requirements is {\it Lynx}. {\it Lynx} is under consideration as a NASA flagship mission going into the 2020 Decadal Survey.  The key features of {\it Lynx} are its angular resolution and sensitivity.  The current concept for {\it Lynx} consists of an X-ray telescope with sub-arcsecond angular resolution comparable to {\it Chandra} ($\sim 0.5^{\prime \prime}$) but with $50 \times$ more collecting area, making it a natural complement to deep, high-resolution radio studies with the ngVLA.

At low redshifts, the ability to spatially and spectrally resolve jet-ISM and radiative wind-ISM interactions in X-rays is critical for testing feedback models. The characteristic temperature of these feedback interactions, which can produce spatially extended line emission indicative of the local conditions, is typically $kT\sim 1\,$keV and is thus uniquely accessible to the X-ray observing regime.  Thus, X-ray observations with {\it Lynx} would be naturally complementary to the ngVLA, which will dramatically improve our ability to find and map radio signatures of feedback.  At high redshifts, excellent spatial resolution is crucial for X-ray studies due to source confusion and the detrimental effect of the X-ray background on faint source detection.

{\it Chandra} is photon-starved in these regimes, such that $\sim$megasecond observations are required to take full advantage of its resolution.  {\it Lynx}, with its significantly greater collecting area compared to {\it Chandra} (and far superior performance at softer energies where AGN feedback traced by shocks is most evident), will allow concurrent surveys of ngVLA targets that will spatially resolve the sites of sub-kpc feedback on scales of tens of pc and microcalorimeter spectral resolution (a few eV, vs. $\sim 100\,$eV for CCDs) which will spectrally resolve critical X-ray transitions unambiguously.  {\it Lynx} will also be the only instrument capable of detecting and matching X-ray counterparts of AGNs identified with the ngVLA at $z>6$, and from heavily obscured AGNs at lower redshifts. Simultaneous radio and X-ray measurements will be critical for investigating the formation and growth of the earliest MBH seeds, since the methods of accretion required for rapid MBH growth in the early universe remain a fundamental unanswered issue.

\section{Summary and Conclusions}
\label{sec:summary}
We have presented an overview of selected areas of research related to AGNs and galaxy evolution that will greatly benefit from the revolutionary capabilities of the ngVLA.  The AGN science topics discussed in our study include jet-ISM feedback associated with radio AGNs that are young and/or compact over a wide range of radio powers and redshifts.  We also emphasized the ability of the ngVLA to pinpoint and study the population of faint AGNs powered by sub-million-solar-mass MBHs, including those hosted by dwarf and low-surface-brightness galaxies.  In addition to probing jet-driven AGN feedback, we also discussed prospects for constraints on radiative-mode feedback through measurements of the SZ decrement in the vicinity of powerful quasars.

To further assess the key ngVLA observational capabilities relevant for AGN science, we produced simulated ngVLA and VLA A-configuration continuum observations of redshifted analogs of three real radio AGNs with diverse jet morphologies spanning a wide range of linear scales (100~kpc, 1~kpc, and 10~pc).  We also reviewed the most significant computational challenges that must be overcome to facilitate the key AGN science topics presented here. 
Finally, we considered opportunities for synergy with multiwavelength observations from current and future telescopes, including the VLA, ALMA, SKA, {\it Lynx}, and {\it JWST}.  Our main conclusions are as follows:

\begin{enumerate}

\item By combining broadband continuum and spectral line observations, the ngVLA will provide exciting new constraints on the physics of jet-driven feedback and the impact it has on the evolution of the host galaxy under a wide variety of conditions.

\item Observations of the low-$J$ CO lines offer the tantalizing possibility of directly probing the cold gas reservoirs of AGNs and the energetic imprint of jet-driven feedback on the ISM of the host.  However, the detectability of these lines will be hampered by the poor brightness temperature sensitivity of the ngVLA on long baselines.  Further studies are needed to address the impact of configuration design on the feasibility of CO observations with the ngVLA.

\item The broad frequency range (1 to 116~GHz) of the ngVLA will facilitate surveys of the radio spectral ages of {\it young} AGNs residing in gas-rich galaxies at low and intermediate redshifts.  This will be highly complementary to studies with the SKA and its pathfinders, which will operate over a lower range of frequencies and be better suited for constraining the radio SEDs and ages of {\it older} and more distant AGNs.

\item We simulated the radio continuum morphologies of redshifted-analogs of three representative radio AGNs with diverse jet properties and intrinsic linear extents ranging from 10~pc to 100~kpc.  For the antenna configuration considered in our simulations with 300 antennas and maximum baselines of $\sim$500~km, the ngVLA is best suited for studies of compact, young radio jets with extents ranging from a few pc to a few kpc. Imaging more extended radio AGNs will not be possible at the maximum resolution of the ngVLA due to surface brightness sensitivity limitations, and will require significant weighting and $uv$-tapering of the data.

\item Radio AGN studies with the ngVLA will require high dynamic range imaging, which will be particularly severe for wide-bandwidth continuum observations using the lower-frequency ngVLA bands.  We thus urge investment in the early development of new calibration and imaging algorithms in the years leading up to the construction of the ngVLA. \\

\end{enumerate}

We emphasize that we have focused our study on the topics of jet-ISM feedback associated with lower-mass, gas-rich host galaxies, AGN feedback at high redshift, and the symbiotic link between the formation and growth of MBHs and the galaxies in which they reside.  The ngVLA will also provide new insights on a wide variety of other topics related to radio AGN science, including the detection of intermediate-mass black holes ($10^2 \leq M_{\mathrm{BH}}/M_{\odot} \leq 10^5$; \citealt{mezcua+18}, and references therein), dual AGNs/recoiling MBHs \citep{komossa+12, blecha+16}, MBH fueling and accretion physics \citep{combes+14, gaspari+15, tremblay+16}, and MBH mass measurements from megamaser (e.g., \citealt{kuo+11, greene+16}) and CO \citep{davis+13, davis+14} dynamics.  

Although we have only considered Stokes~$I$ continuum simulations for a handful of sources, full-polarization and spectral line simulations are clearly warranted, particularly to address the need for the new tapering and weighting algorithms that will be necessary for the detection of the low-$J$ CO lines.  

\section*{Acknowledgments}
We thank the anonymous referee for carefully reading our manuscript and providing us with constructive comments that have improved the quality of this work.  
The National Radio Astronomy Observatory is a facility of the National Science Foundation operated under cooperative agreement by Associated Universities, Inc.  This work was funded in part by the grant associated with {\it Spitzer} proposal 11086.  This study includes observations made with the NASA/ESA {\it Hubble} Space Telescope, and obtained from the {\it Hubble} Legacy Archive, which is a collaboration between the Space Telescope Science Institute (STScI/NASA), the Space Telescope European Coordinating Facility (ST-ECF/ESA) and the Canadian Astronomy Data Centre (CADC/NRC/CSA).  W.R. is supported by JSPS KAKENHI Grant Number JP15K17604, Thailand Research Fund/Office of the Higher Education Commission Grant Number MRG6080294, and Chulalongkorn University's CUniverse.

\vspace{5mm}

\bibliographystyle{apj}
\bibliography{ngVLA_AGNs_white_paper_v6_arxiv}

\end{document}